\pdfoutput=1 

\documentclass[prd,aps,preprint,amsmath,nofootinbib,amssymb,eqsecnum,showkeys,tightenlines]{revtex4-1}

\usepackage{amsmath, latexsym, amssymb, hyperref, graphicx, color, multirow}
\usepackage[table]{xcolor}
\usepackage{tabularx}
\usepackage[T1]{fontenc}
\usepackage[capitalize]{cleveref}
\definecolor{nicered}{rgb}{.7,.1,.1}
\definecolor{nicegreen}{rgb}{.1,.5,.1}
\definecolor{darkblue}{rgb}{0,0,.5}
\hypersetup{colorlinks, citecolor=nicegreen,linkcolor=nicered, urlcolor=darkblue}
\usepackage{multirow}
\usepackage{slashed}
\numberwithin{equation}{section}

\begin{document}

\title{Characterizing Higgs portal dark matter models at the ILC}

\author{Teruki Kamon}
\email{kamon@physics.tamu.edu}
\affiliation{Mitchell Institute for Fundamental Physics and Astronomy, Department of Physics and Astronomy, Texas A\&M University, College Station, TX 77843-4242, USA}

\author{P. Ko}
\email{pko@kias.re.kr}
\affiliation{School of Physics, Korea Institute for Advanced Study, Seoul 130-722, Korea}

\author{Jinmian Li}
\email{jmli@kias.re.kr}
\affiliation{School of Physics, Korea Institute for Advanced Study, Seoul 130-722, Korea}

\begin{abstract}
\noindent
We study the Dark Matter (DM) discovery prospect and its spin discrimination in the theoretical framework of 
gauge invariant and renormalizable Higgs portal DM models at the ILC with $\sqrt{s} = 500$ GeV. 
In such models, the DM pair is produced in association with a $Z$ boson. In case the singlet scalar DM, 
the mediator is just the SM Higgs boson, whereas for the fermion or vector DM there is an additional singlet 
scalar mediator that mixes with the SM Higgs boson, which produces significant observable differences. 
After careful investigation of the signal and backgrounds both at parton level and at detector level, we find 
the signal with hadronically decaying $Z$ boson provides a better search sensitivity than the signal with leptonically decaying $Z$ boson. 
Taking the fermion DM model as a benchmark scenario, when the DM-mediator coupling $g_\chi$ is relatively small, the DM signals are discoverable only for benchmark points with relatively light scalar mediator $H_2$. The spin discriminating from scalar DM is always promising while it is difficult to discriminate from vector DM. 
As for $g_\chi$ approaching the perturbative limit, benchmark points with the mediator $H_2$ in the full mass 
region of interest are discoverable. The spin discriminating from both the scalar and fermion DM are quite promising.
\end{abstract}

\maketitle

\section{Introduction}\label{sec:intro}

Since the existence of Dark Matter (DM) is confirmed by many astrophysical observations~\cite{Ade:2015xua}, identifying the properties such as their masses and spins and couplings of the DM is one of the most important 
tasks of particle physics. 
The most often considered thermal DM candidate is the weakly interacting massive particle (WIMP), 
which has the mass around $\mathcal{O}(100)$ GeV and interacts with Standard Model (SM) particles 
via the electroweak force. Thus it can be produced directly in collider experiments in principle. 
The DM signal at colliders can be probed as the momentum imbalance at the detector if it is produced with 
recoiling against visible objects. 
Probing the DM signals at colliders could elucidate the particle physics properties of DM without suffering from astrophysical uncertainties thus becomes one of the main object of the current and future colliders. 

There are three theoretical frameworks~\cite{Kahlhoefer:2017dnp} that are used for describing the DM phenomena at the colliders, each has its own advantages and limitations:
\begin{itemize}
\item The DM effective field theory (EFT)~\cite{Goodman:2010yf,Goodman:2010ku,Duch:2014xda} is the low energy approximation of a renormalizable theory after integrating out the heavy particle that mediates the DM-SM particles interactions. The number of free parameters in the EFT is minimal, i.e. only two parameters are relevant for each 
operator, the coefficient of the effective operator and the DM mass. 
However, the EFT descriptions of DM interactions are valid only when momentum transfer is much smaller than the mass of the mediator such as in DM direct detection experiments. While in the collider experiments, where the momentum transfers can be quite high, the kinematic distributions that are predicted by the underlying UV completion are not correctly captured by the EFT~\cite{Buchmueller:2013dya,Busoni:2013lha,Busoni:2014sya,Busoni:2014haa,Pobbe:2017wrj}, especially in the region with light mediator or heavy DM. 

\item In the DM simplified model~\cite{Abdallah:2014hon,Abdallah:2015ter,Abercrombie:2015wmb}, the DM is neutral under the SM gauge group and interacts with the SM particles via the portal of a single particle. Models in this class usually contain 5 free parameters: DM mass $m_\chi$, mediator mass $m_{\text{med}}$, DM-mediator coupling $g_\chi$, SM particle-mediator coupling $g_{\text{SM}}$ and the mediator decay width $\Gamma_{\text{med}}$. Considering mediators of different masses makes it possible to consider different kinematic distributions that cannot be mapped onto effective operators, thus providing a more general framework for describing the DM phenomena. However, simplified DM models with a single scalar mediator often violate the SM gauge symmetry 
\cite{Baek:2015lna,Ko:2016ybp}
and perturbative unitarity, thus may become invalid for describing some sort of UV-complete models. 

\item There are growing interests in second generation simplified DM model that respects the SM gauge symmetry 
and renormalizability~\cite{Baek:2011aa,Baek:2012uj,Ko:2016ybp,Baek:2015lna,Kahlhoefer:2015bea,Englert:2016joy,Bell:2015sza,Bauer:2016gys,Ko:2016zxg,Baek:2012se,Bell:2016ekl,Baek:2017vzd}. Among them, the simplest ones are singlet DM extension of SM with Higgs portal. In these models, depending on the DM spin, the gauge invariant DM-SM interactions may require at least two mediators. 
Even though in the parameter region where only the contribution of one mediator dominates, the prediction of this model coincide with that of the simplified model with a single mediator. We have shown in Ref.~\cite{Ko:2016ybp} that the interference effect between the two mediators can affect the exclusion bounds considerably in some parameter space \footnote{See also Ref.~\cite{Baek:2015lna} for earlier study.}. 
While the models in this class give more realistic predictions regarding a UV completion, there will be ad hoc constraints from many experiments, which may be quite specific and only applicable to certain UV completed models. 
For example, for singlet fermion DM extension of SM, the constraints and the prediction of the model with $h_{\text{SM}}$ + singlet scalar portal is quite different from those of the model with two-Higgs-doublet portal. 
\end{itemize}

All the above frameworks have been widely used in studying DM phenomenology at colliders. 
Each case contains quite a lot of possible operators/models that describe the nature of DM and its couplings. 
If any excess is observed within a given theoretical framework, it will be important to ask which operator/model 
provides the best description, i.e. characterize the DM properties. There are several studies~\cite{Cotta:2012nj,Crivellin:2015wva,Belyaev:2016pxe} devoted to distinguishing the DM EFT operators and its spin at the LHC. 
In the framework of DM simplified model with single mediator, many current works~\cite{Haisch:2013fla,Dolan:2016qvg,Haisch:2016gry,Haisch:2013ata,Buckley:2014fba,Harris:2014hga,Haisch:2015ioa,Backovic:2015soa,Buckley:2015ctj} are mainly focused on distinguishing the spin of the mediator and identifying the coupling forms between the 
mediator and SM particles. Because here the DM are dominantly produced by the decay of the on-shell mediator, those visible final states does not carry any information of the DM nature. 

In this work, based on the gauge invariant and renormalizable DM models with Higgs portal, we will study the fermion DM (FDM) discovery prospects and its spin discriminations against scalar DM (SDM) and vector DM (VDM) at the ILC.  A very preliminary study along the similar direction has been given by one of us in Ref.~\cite{Ko:2016xwd}, where the detector effects were completely ignored in discussions of the DM discovery and only qualitative arguments were given regarding the spin discrimination. 
By curing these two problems, we find the hadronic channel of DM production provides a better sensitivity for DM discovery than the leptonic channel. Taking the FDM model as a reference model, the FDM with the coupling in a wide range can be discovered in the hadronic channel when the second mediator is relatively light. In this region, the spin discriminating from SDM is always quite promising, because the SDM model is intrinsically different from the FDM model with only one mediator being introduced. However, the spin discriminating from VDM is much more difficult, which becomes possible only in the region where the coupling between the DM and the second mediator is approaching the perturbative limit. 

This paper is organized as follows. In Sec.~\ref{sec:model}, we briefly describe the renormalizable and gauge 
invariant Higgs portal DM models for scalar, fermion and vector DM. Their complete Lagrangians as well as the interaction Lagrangians that are relevant to the DM search at collider are provided. Sec.~\ref{sec:g3had} details the analysis for the DM discovery and the strategy for the DM spin discrimination based on a benchmark scenario. Similar methods are then applied to the leptonic channel of the benchmark scenario and the hadronic channel with different couplings in Sec.~\ref{sec:g3lep} and Sec.~\ref{sec:g110had}, respectively.   Then we conclude the work in Sec.~\ref{sec:concl}.

\section{Higgs portal DM models}
\label{sec:model}
In this section, we define the Higgs portal DM models with SM gauge invariance and renormalizability, 
where the DMs are scalar, fermion and vector particle, respectively. 

The SDM model can be constructed by simply introducing a new scalar $S$ in addition to the SM~\cite{SILVEIRA1985136,Burgess:2000yq,Queiroz:2014pra}
\begin{equation} \label{eq:lgsdm}
{\cal L}_{\rm SDM} = \frac{1}{2} \partial_\mu S \partial^\mu S 
- \frac{1}{2} m_{0}^2 S^2 - \lambda_{HS} H^\dagger H S^2  
- \frac{\lambda_S}{4 !} S^4,
\end{equation}
where $H$ is the SM Higgs doublet and $S$ is assumed to be odd under a $Z_2$ symmetry and thus becomes a DM candidate. After the electroweak (EW) symmetry breaking $H \to (0, (v_h+h)/\sqrt{2})^T$ and assuming $\langle S \rangle =0$, we can write down the interaction Lagrangian for DM production at the ILC as
\begin{align}
\mathcal{L}^{\text{int}}_{\text{SDM}} =  - h~ \left( \frac{2 m^2_W}{v_h} W^+_\mu W^{- \mu} + \frac{m^2_Z}{v_h} Z_\mu Z^\mu \right)   - \lambda_{HS} v_h~ h S^2.  \label{eq:lsdm}
\end{align}
In this model, the DM can only be pair produced through the SM Higgs ($h$) mediation. 

The simplest Higgs portal singlet FDM model with SM gauge invariance and renormalizability 
contains a SM singlet Dirac fermion DM $\chi$ and a real singlet scalar mediator $S$ 
\footnote{Here the singlet scalar $S$ is different from the singlet scalar DM defined in Eq.~(\ref{eq:lgsdm}), 
although we use the same notation.  In the FDM case, there is no $Z_2$ symmetry ($S\rightarrow -S$) 
so that $S$ cannot be a DM candidate, and $S$ is a messenger between the dark
sector and the SM sector through the Yukawa coupling ($y_\chi$-term) in Eq.~(\ref{eq:lgfdm}).} 
in addition to the SM particles~\cite{Baek:2011aa,Baek:2012uj}:
\begin{eqnarray} \label{eq:lgfdm}
{\cal L}_{\rm FDM} &=& \overline{\chi} \left( i \slashed{\partial} - m_\chi
 - y_\chi S \right) \chi + \frac{1}{2} \partial_\mu S \partial^\mu S  -
 \frac{1}{2} m_0^2 S^2
\\
& -& \lambda_{HS} H^\dagger H S^2 - \mu_{HS} S H^\dagger H -
 \mu_0^3 S
 - \frac{\mu_S}{3 !} S^3 - \frac{\lambda_S}{4 !} S^4,
\nonumber 
\end{eqnarray}
where the singlet scalar $S$ can not have direct renormalizable couplings to the SM particles due to the SM 
gauge symmetry and the singlet Dirac fermion $\chi$ is assumed to be odd under a $Z_2$ dark parity 
$\chi \rightarrow - \chi$.  
When both scalar fields $H$ and $S$ develop nonzero vacuum expectation values (VEV), $v_h$ and $v_s$, so that
\begin{equation}
H = 
\left( \begin{array}{c}
  G^+ \\
  \frac{1}{\sqrt{2}} (v_h + h + iG^0)
  \end{array} \right) , ~~~~~ S= v_s +s~,~
\end{equation}
the two scalar fields mix
\begin{equation}
\left(  \begin{array} {c}
  h \\
  s
       \end{array} \right) = 
       \left(  \begin{array}{c c}
         \cos \alpha & \sin \alpha \\
         -\sin \alpha & \cos \alpha
         \end{array}  \right) 
         \left(
           \begin{array}{c}
           H_1 \\
           H_2
           \end{array}
         \right)~,~ \label{eq:hmix}
\end{equation} 
giving $H_1$ and $H_2$ fields in mass eigenstate. The mixing angle can be expressed in terms of parameters in scalar potential
\begin{align}
\tan 2\alpha = - \frac{2 \lambda_{HS} v_s v_h + 2 \mu_{HS} v_h}{ 2 \lambda_S v^2_s - \frac{\mu_0^3}{v_s} - \mu_S v_s - \frac{\mu_{HS} v^2_h}{2 v_s} - 2\lambda_{H} v^2_h  } ~.~
\end{align} 

The interaction Lagrangian of interest can be written in the mass eigenstates as
\begin{align}
\mathcal{L}^{\text{int}}_{\text{FDM}} &= - \left(H_1 \cos \alpha + H_2 \sin \alpha \right) \left( \sum_f \frac{m_f}{v_h} \bar{f} f 
- \frac{2m^2_W}{v_h} W^+_\mu W^{-\mu} - \frac{m_Z^2}{v_h} Z_\mu Z^\mu  \right) \nonumber \\ 
     & + g_\chi \left(H_1 \sin \alpha - H_2 \cos \alpha \right) ~ \bar{\chi} \chi ~.~ \label{eq:lfdm}
\end{align} 
In contrast to the SDM model, there are two scalar bosons that mediate the DM production in the fermion DM model. The interference effects between two mediators can lead to interesting applications to DM searches at colliders~\cite{Baek:2015lna,Ko:2016ybp}. 
If the $H_1$ is assumed to be the 125 GeV Higgs boson~\cite{Aad:2012tfa,Chatrchyan:2012xdj} with its measured strengths~\cite{Khachatryan:2014jba,Aad:2015gba}, the mixing angle should be small, $\sin \alpha \lesssim 0.4$~\cite{Robens:2015gla,Cheung:2015dta,Dupuis:2016fda}. 
 
As for constructing a renormalizable and gauge invariant model for vector (VDM), we need to introduce an 
abelian dark gauge group $U(1)_{X}$ and a dark Higgs field $\Phi$~\cite{Farzan:2012hh,Baek:2012se}:
\begin{equation}
{\cal L}_{\text{VDM}} = -\frac{1}{4} V_{\mu\nu} V^{\mu\nu} + D_\mu \Phi^\dagger D^\mu \Phi 
- \lambda_\Phi \left( \Phi^\dagger \Phi - \frac{v_\phi^2}{2} \right)^2 
- \lambda_{H\Phi}   \left( H^\dagger H - \frac{v_h^2}{2} \right) 
\left( \Phi^\dagger \Phi - \frac{v_\phi^2}{2} \right), 
\end{equation}
where the VEV of $\Phi = \frac{1}{\sqrt{2}} \left(v_\phi +\phi \right)$ will provide mass to the vector DM $V_\mu$. 
The convariant derivative is defined as $D_\mu \Phi = \left( \partial_\mu + i g_V Q_\Phi V_\mu \right) \Phi$ where the $U(1)_X$ charge of $\Phi$ will be taken as $Q_\Phi=1$ throughout the paper. 
In this model, a $Z_2$ symmetry ($V_\mu \to - V_\mu$) and charge conjugation symmetry have been imposed 
by hand, thereby forbidding the kinetic mixing between $V_\mu$ and the SM $U(1)_Y$ gauge boson and making  the vector boson $V_\mu$ stable. It can also be implemented by some unbroken local dark gauge symmetry 
as proposed in Ref.~\cite{Baek:2013dwa}.

Similarly to the FDM model with Higgs portal, there are two scalar mass eigenstates ($H_{1/2}$) that are 
originated from the mixing of SM Higgs $h$ and dark Higgs $\phi$, with the mixing angle given by 
\begin{align}
\tan  2 \alpha = \frac{ \lambda_{H\Phi} v_h v_\phi}{\lambda_\Phi v^2_\phi - \lambda_H v^2_h}.
\end{align} 
Then, the interaction Lagrangian that is relevant to the collider study can be written as 
\begin{align}
\mathcal{L}^{\text{int}}_{\text{VDM}} &= - \left(H_1 \cos \alpha + H_2 \sin \alpha \right) \left( \sum_f \frac{m_f}{v_h} \bar{f} f 
- \frac{2m^2_W}{v_h} W^+_\mu W^{-\mu} - \frac{m_Z^2}{v_h} Z_\mu Z^\mu  \right) \nonumber \\ 
     & - \frac{1}{2} g_V m_V \left(H_1 \sin \alpha - H_2 \cos \alpha \right) ~V_\mu V^\mu ~.~ \label{eq:lvdm}
\end{align}

So far we have derived the relevant interaction Lagrangians for scalar, fermion and vector DMs with Higgs portal 
in  Eqs.~(\ref{eq:lsdm}),~(\ref{eq:lfdm}),~(\ref{eq:lvdm}) respectively. Note that there is only one scalar mediator 
($h$) in the scalar DM model, while there are two scalar mediators ($H_{1/2}$) in fermion and vector DM models. 
The difference in the number of mediators can lead to quite different kinematic distributions, which can be used to discriminate scalar DM model against fermion/vector DM models. 
On the other hand, distinguishing fermion DM models from vector DM models is more involved. 
First of all, if the DM production is dominated by on-shell $H_{1/2}$ production with subsequent invisible decay, it will be impossible to observe any differences in the final state distribution. The spin discrimination between fermion and vector DM is possible only if the off-shell contributions become important.  Then, given the same decay width of $H_{1/2}$, the fermion and vector DM model will predict different DM production rate as well as final state kinematics. 

\section{A benchmark study}
\label{sec:g3had}
At the ILC, the Higgs portal DM is dominantly produced through the Higgs-strahlung process
\begin{align}
e^+ e^- \to Z H_{1/2} \left( \to DD \right), \label{sigproc}
\end{align}
where $D = S, \chi , V_\mu$ for scalar, fermion and vector DM, respectively. The $Z$ boson can decay either leptonically or hadronically.  We will show later that the leptonic mode which is suffering from branching ratio suppression is less sensitive than the hadronic mode. In this section, we will focus on the discovery prospect of the hadronic mode of fermion DM and discuss its spin discrimination against vector/scalar DM. 
Note that for scalar DM, only one mediator $H_1 = h_{\text{SM}}$ is introduced. 

In order to guarantee sufficient DM production rate at colliders while consistent with current measurements, 
the relevant parameters for the fermion DM production are chosen as 
\begin{align}
g_\chi=3, ~ \sin \alpha =0.3, ~ m_{\chi} =80~ \text{GeV}. 
\end{align}
Four benchmark points with different $m_{H_2}=$ (200, 300, 400, 500) GeV will be studied, which are denoted 
as FDM200, FDM300, FDM400 and FDM500, respectively. 
For each benchmark point, we assume that the decay width for heavier scalar $H_2$ into the $H_1$ pair is 
negligible \footnote{This depends on a new parameter from the scalar potential, and we ignore it in order to simplify the discussion. If there is $H_2 \to H_1 H_1$ decay, our DM production cross section will be suppressed by branching ratio. Meanwhile the total decay width of $H_2$ will be broadened allowing more off-shell contributions~\cite{Ko:2016ybp}. }.  Then we can express the minimal decay width of $H_2$ as
\begin{align}
\Gamma^{\text{FDM}}_{\min}(H_2) &= \Gamma \left(H_2 \to \chi \chi \right) + \Gamma\left( H_2 \to WW/ZZ \right) + \Gamma\left( H_2 \to ff \right)  \nonumber \\
  &= \cos^2 \alpha \cdot g^2_\chi \frac{m_{H_2}}{8 \pi} \left( 1- \frac{4 m^2_\chi}{m^2_{H_2}} \right)^{3/2} + {\sin^2 \alpha \cdot \frac{G_\mu m^3_{H_2}}{16\sqrt{2} \pi} \delta_V \sqrt{1-4 \frac{m_V^2}{m^2_{H_2}}} \left( 1-4 \frac{m_V^2}{m^2_{H_2}} +12 \frac{m_V^4}{m^4_{H_2}} \right)} \nonumber \\
     &+ \sin^2 \alpha \cdot \left( \frac{m_f}{v} \right)^2 \frac{3 m_{H_2}}{8 \pi} \left(1-\frac{4 m^2_f}{m^2_{H_2}} \right)^{3/2}~,  \label{adecay}
\end{align}
where $f$ is the SM fermion, $V=Z, W$ and $\delta_V=1(2)$ for $Z(W^{\pm})$. 

To study the spin discrimination, the parameters for the vector DM production are chosen accordingly: 
\begin{align}
\sin \alpha =0.3, ~ m_{V} =80~ \text{GeV}
\end{align}
and the $g_V$ is chosen such that the total decay width of $H_2$ is the same with that in the fermion DM case, since one can rely on other method to measure the total decay width of $H_2$. We give the total decay widths of $H_2$ for four benchmark points in FDM model as well as the corresponding $g_V$ of VDM model in Tab~\ref{FVgg}.  Due to the different dependencies of the $\Gamma_{H_2}$ on the $m_{H_2}$ in FDM and VDM models, the $g_V$ can be quite different from the $g_\chi$ (=3).  In VDM model, a heavier $H_2$ requires a smaller $g_V$ to maintain the decay width of $H_2$ being the same with that in FDM. 
The decay width of $H_2$ here is similar to the Eq.~(\ref{adecay}) with the first term $\Gamma(H_2 \to \chi \chi)$ replaced by $\Gamma(H_2 \to V V)$. 

\begin{table}[htb]
\begin{center}
\begin{tabular}{|c||c|c|c|c|c|} \hline
$m_{H_2}$ [GeV] & 200 & 300 & 400 & 500\\ \hline
$\Gamma_{\min} (H_2)$ [GeV] & 14.2 & 60.1 & 103.0 & 144.5 \\  \hline
$g_V$ & 3.53 & 3.07 & 2.37 & 1.91 \\ \hline
\end{tabular}
\caption{\label{FVgg} First two rows are the masses and decay widths of the $H_2$ for four benchmark points in FDM model. The last row gives the $g_V$ in VDM model which produce the $\Gamma_{\min} (H_2)$ in the second row. }
\end{center}
\end{table}

The scalar DM model is much simpler, since there are only two parameters: $m_S$ and $\lambda_{HS}$.  In studying the spin discrimination, we will fix $m_S = 80$ GeV and take appropriate $\lambda_{HS}$ such that the number of signal events after all selections are kept the same as that of each benchmark point of the FDM model. 
However, changing the $\lambda_{HS}$ can only lead to total rescaling of the cross section and will not affect the differential distribution of kinematic variables. In the following, we will fix $\lambda_{HS}$ when discussing the kinematic shapes without loss of generality.  

\begin{figure}[htbp]
\begin{center}
\includegraphics[width=0.3\textwidth]{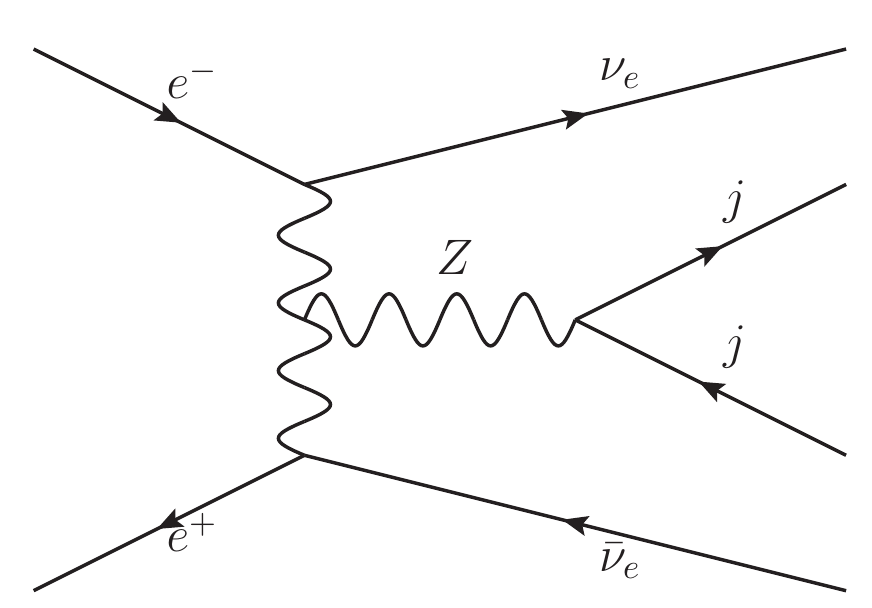}
\includegraphics[width=0.3\textwidth]{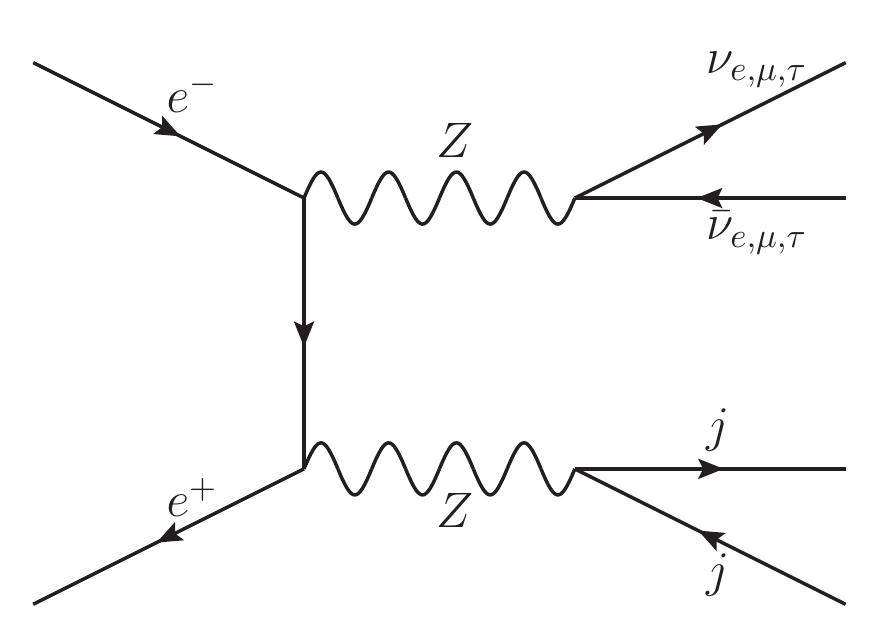} 
\includegraphics[width=0.3\textwidth]{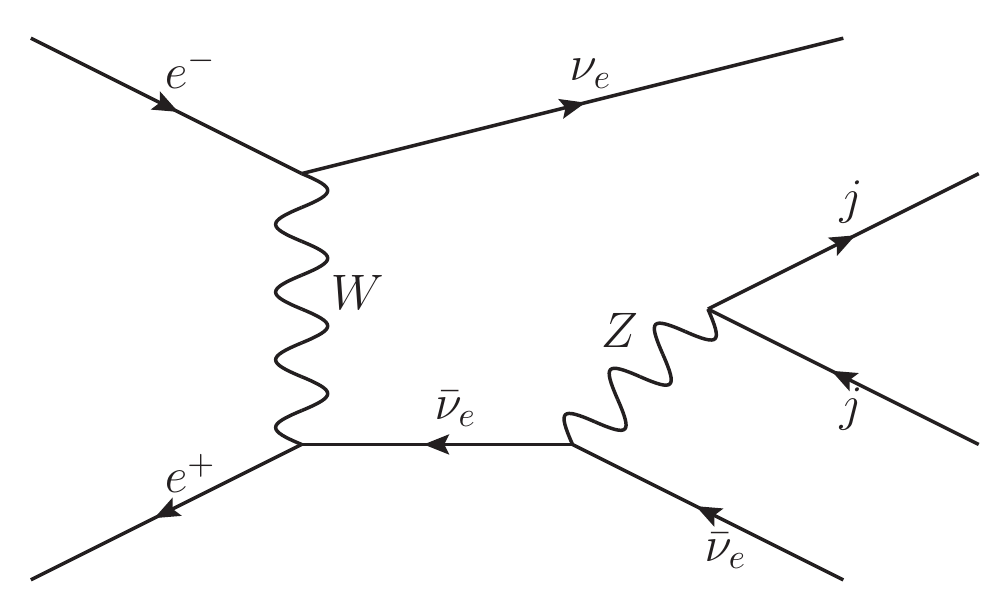}
\end{center}
\caption{\label{bgs} Dominant background processes for hadronic mode of our signal.}
\end{figure}

\begin{figure}[htbp]
\begin{center}
\includegraphics[width=0.47\textwidth]{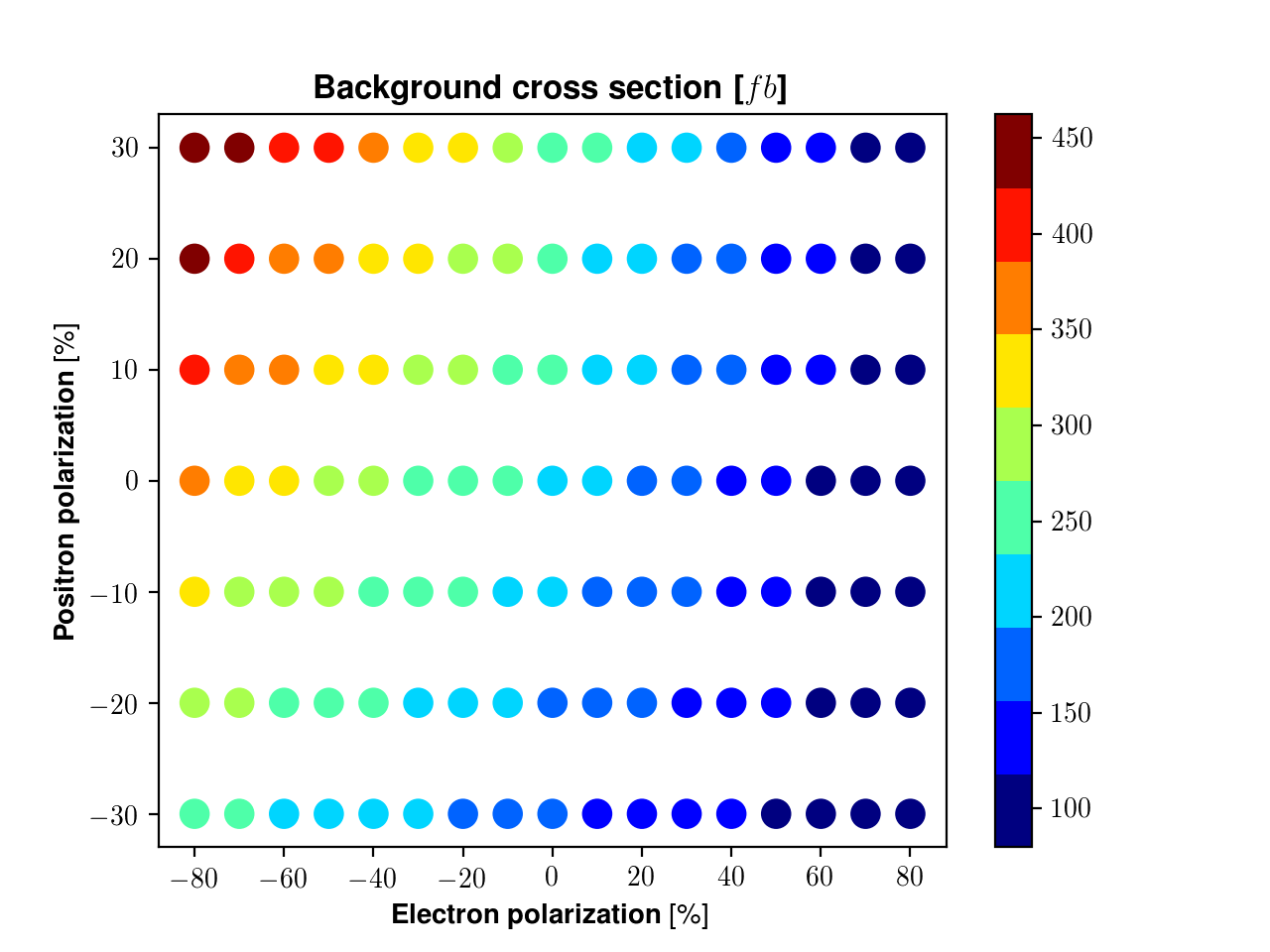}
\includegraphics[width=0.47\textwidth]{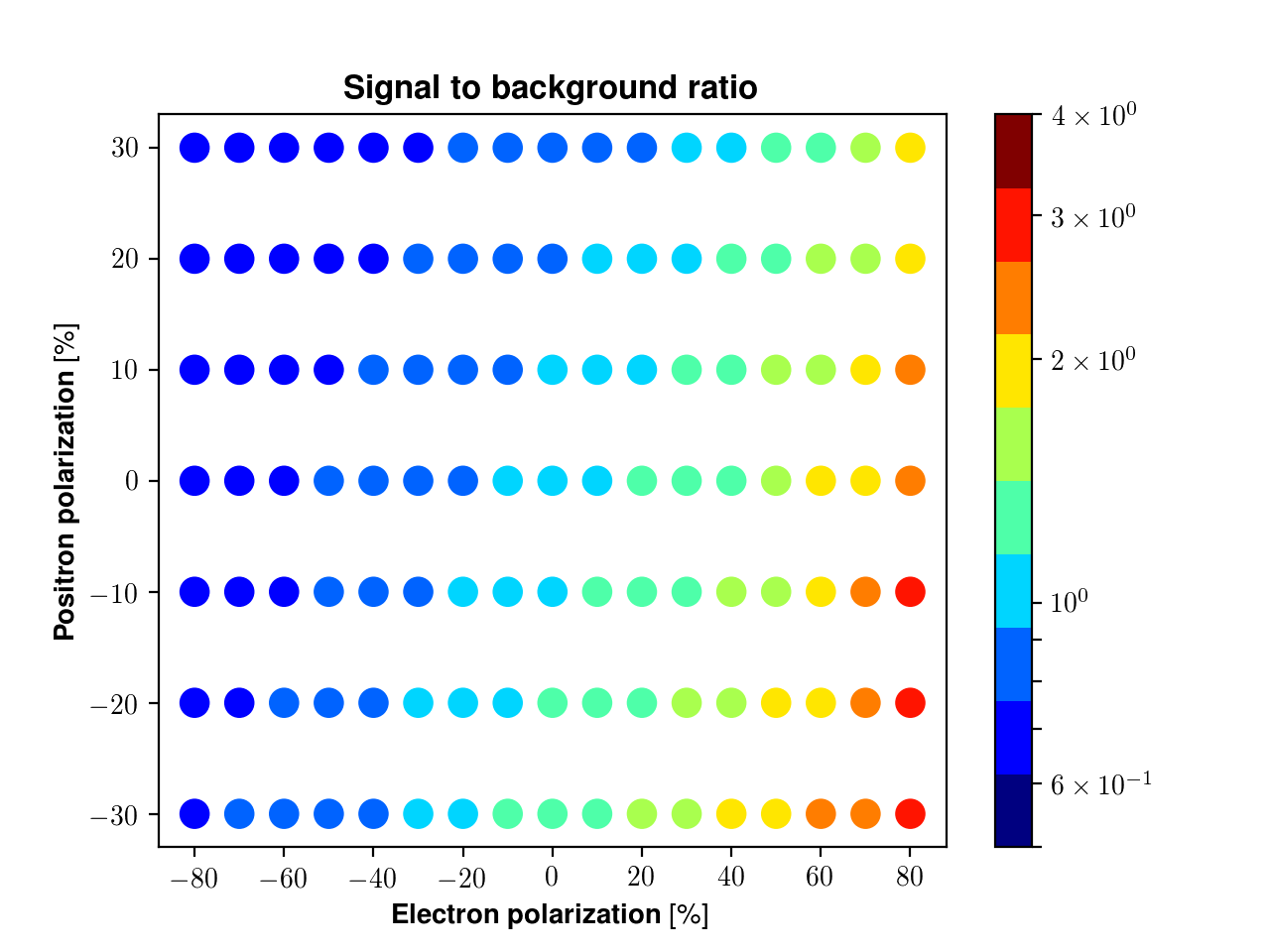}
\end{center}
\caption{\label{fig:polar} The background cross section (left) and signal to background ratio (right) with varying electron and position beam polarization at the $\sqrt{s}=500$ GeV. In the right panel, the benchmark point of FDM200 has been taken as signal for illustration. The signal to background ratios have been normalized at $(P_{e^-} , P_{e^+}) = (0,0)$. }
\end{figure}

The SM processes with any species of neutrino in the final state could mimic the DM signal. 
The dominant SM background processes to Eq.~(\ref{sigproc}) are shown in Fig.~\ref{bgs}. Among them, the first and the second diagram (including three species of neutrino) give similar amount of contributions, while the third one is slightly smaller. 
At the ILC with $\sqrt{s}=500$ GeV and unpolarized beams, the total production cross section including the interference effects between different diagrams is 219 fb. Since the left and right handed fermions have different electroweak charges, the background cross section, especially the contribution from vector boson fusion (VBF) process, strongly depends on the beam polarization. The ILC will be able to provide highly polarized electron beam (80\%) and moderately polarized positron beam (30\%)~\cite{Asner:2013psa}. The background cross sections with respect to the varying beams polarizations are plotted in the left panel of Fig.~\ref{fig:polar}, where we have used the positive sign for right handed polarization and negative sign for left handed polarization. We can see that the background cross section is largest for electron polarization $P_{e^-}=-80\%$ and positron polarization $P_{e^+}=30\%$, while it is smallest for $(P_{e^-} , P_{e^+})= (80\%,-30\%)$. Meanwhile, the cross sections of signal processes also mildly depend on the beam polarization. Taking the benchmark point of FDM200 for illustration, the signal to background ratio with respect to the varying beams polarizations are given in the right panel of Fig.~\ref{fig:polar}, where the values have been normalized to unit at $(P_{e^-} , P_{e^+}) = (0,0)$. It can be seen that the signal to background ratio can be either reduce by a factor of $0.7$ or enhanced by a factor of $\sim 3$ comparing to the value at $(P_{e^-} , P_{e^+}) = (0,0)$.
Although polarized beams improve the sensitivity, we report the results with the unpolarized beam in this work. 

In this work, the cross sections and events for signal and background are generated by MadGraph5\_aMC@NLO\_v2.4.3~\cite{Alwall:2014hca}. The Pythia6~\cite{Sjostrand:2006za} is used for parton showering and hadronization. The final state jets are clustered using the Fastjet~\cite{Cacciari:2011ma}. We also include the detector effects by using Delphes\_v3.4.1~\cite{deFavereau:2013fsa} with input of ILD card~\cite{Behnke:2013lya}. The track momentum and calorimeter energy resolutions of the card are listed in Tab.~\ref{resolu}. It should be noted that a more realistic detector simulation should also consider the energy spectra of income beams, the effect of which is neglected in our simulation. 

\begin{table}[htb]
\begin{center}
\begin{tabular}{|c||c|} \hline
\multirow{2}{*}{Track momentum}& $10^{-5} \oplus \frac{0.001}{p_T}$ , for $\left| \eta \right| \le 1.0$ \\
    & $10^{-4} \oplus \frac{0.01}{p_T}$ , for $\left| \eta \right| \in \left( 1 ,2.4 \right]$  \\ \hline
Electromagnetic calorimeter  & $1\% \oplus \frac{0.15}{\sqrt{E}} $ , for $\left|\eta \right| \le 3.0$ \\ \hline
Hadronic calorimeter &  $1.5\% \oplus \frac{0.5}{\sqrt{E}} $ , for $\left|\eta \right| \le 3.0$  \\\hline
\end{tabular}
\caption{\label{resolu} The resolutions for track momentum ($\sigma_{1/p_T}$), electromagnetic calorimeter ($\sigma_{1/E}$) and hadronic calorimeter ($\sigma_{1/E}$).  }
\end{center}
\end{table}

\subsection{Features of DM spin}

For our signal processes at the ILC, the 4-momentum of the DM pair system can be solved as 
\begin{align}
P^\mu_{DD} &=P^\mu_{e^+} + P^{\mu}_{e^-} - P^\mu_Z \nonumber \\
 &= \left( \sqrt{s}- E_Z, - \vec{p}_Z \right), 
\end{align}
where the $\sqrt{s}$ is the collision energy and $E_Z$ ($\vec{p}_Z$) is the energy (momentum) of the $Z$ boson. 
Therefore the invariant mass of the DM system is an observable at the ILC: 
\begin{align}
m^2_{DD} = s +m^2_Z - 2 E_Z \sqrt{s} \ . \label{eq:mdd}
\end{align}
The differential cross section with respect to $m^2_{DD}$ for scalar, fermion and vector DM production have been calculated in Ref.~\cite{Ko:2016xwd}. It can be factorized as an off-shell mediator production and decay: 
\begin{align}
\frac{d \sigma_D}{ d t} =  \frac{1}{2 \pi} \sigma_{h^* Z} \left(s, t \right) \cdot G_D \left( t \right), \label{diffs}
\end{align}
where $t \equiv m^2_{DD}$ and $D= S,\chi ,V$ for scalar, fermion, vector DM respectively. The off-shell mediator production cross section 
\begin{align}
\sigma_{h^* Z} \left(s, t \right) =\mathcal{P}_{ee}  \frac{1}{6 s} \frac{m^4_Z}{v_h^4} \left| \frac{s}{s- m^2_Z + i m_Z \Gamma_Z} \right|^2 \frac{\hat{\beta}}{8 \pi} \left( \hat{\beta}^2 + \frac{12 m_Z^2}{s} \right) \label{shz}
\end{align}
is universal for all DM spins. In above equation, $\mathcal{P}_{ee} = \left(-\frac{1}{2} + 2 \sin \theta_W \right)^2+ \left(-\frac{1}{2} \right)^2 $ with $\theta_W$ being the weak mixing angle is the averaged spin factor for initial electron and positron; $\hat{\beta} = \lambda^{1/2} \left(1, m^2_Z /s, t/s \right)$ with $\lambda \left(a,b,c \right) = a^2+b^2+c^2 -2 \left(ab+bc+ca \right)$.
The $G_D \left(t \right)$ in Eq.~(\ref{diffs}) which is different from spin to spin shows the spin dependent behaviour of the differential cross section:
\begin{align}
G_S(t) & = \frac{\beta_S}{8 \pi} \cdot \left| \frac{\lambda_{HS} v_h}{t - m^2_h + i m_h \Gamma_h} \right|^2, \\
G_\chi (t) &= \frac{\beta_\chi^3}{8 \pi} 2 g_\chi t \cdot \left| \frac{1}{t- m_{H_1}^2 + i m_{H_1} \Gamma_{H_1} } - \frac{1}{t- m_{H_2}^2 + i m_{H_2} \Gamma_{H_2}} \right| ^2, \\
G_V(t) &= \frac{\beta_V}{16\pi} \frac{g_V^2 t^2}{4 m^2_V} \left(1- \frac{4 m^2_V}{t} + \frac{12 m^4_V}{t^2} \right) \cdot \left|\frac{1}{t- m_{H_1}^2 + i m_{H_1} \Gamma_{H_1} } - \frac{1}{t- m_{H_2}^2 + i m_{H_2} \Gamma_{H_2}}\right| ^2 ,
\end{align}
where $\beta_{S/\chi/V} = \sqrt{ \left(1- 4m^2_{S/\chi/V} / t \right)}$.

\begin{figure}[htbp]
\begin{center}
\includegraphics[width=0.7\textwidth]{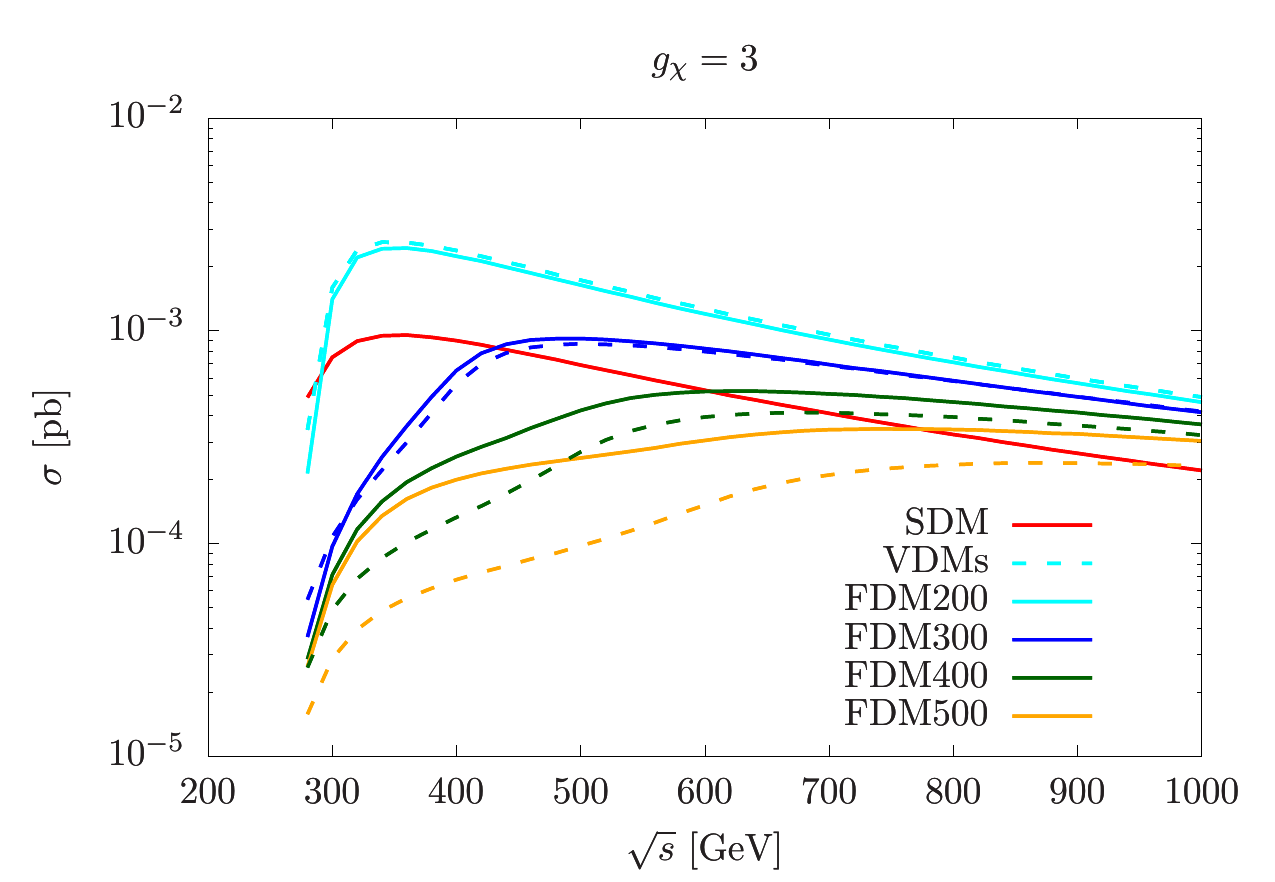}
\end{center}
\caption{\label{sigmas} Production cross section of $e^+ e^- \to Z \left(\to j j \right) DD$ for benchmark points in FDM and VDM models with varying collision energy $\sqrt{s}$ as well as that in SDM with $\lambda_{HS}=1$. The meanings of lines with different colors are indicated in the legend. The dashed lines correspond to the benchmark points in VDM model which have the same $H_2$ mass with the points in FDM model that is shown by the solid line with the same color. }
\end{figure}

We can see from above that different DM spins can lead to different collision energy $\sqrt{s}$ dependence of the production cross sections and different distributions of the DM pair invariant mass $m_{DD}$. 
Especially the threshold behaviors ($t \gtrsim 4 m_{DD}^2$) or the large-$t$ bahaviors 
clearly depend on the DM spin.
In Fig.~\ref{sigmas}, we show the DM total production cross section in SDM, FDM and VDM models by integrating over $t$ in Eq.~(\ref{diffs}). The cross sections of benchmark points in FDM and VDM increase faster than that in SDM, due to the contributions from the second mediator. Comparing FDM and VDM, we can find that the VDM has slightly larger cross section than FDM when the $m_{H_2} \lesssim 200$ GeV, while it can have much smaller cross section for heavy $H_2$. The differences are largest when the collision energy is relatively small $\sqrt{s} \sim [400,500]$ GeV. In the following discussion, we will study the collider phenomenology with fixed $\sqrt{s}=500$ GeV, so that FDM and VDM may possibly be distinguished by their production rate directly.  

\begin{figure}[htbp]
\begin{center}
\includegraphics[width=0.4\textwidth]{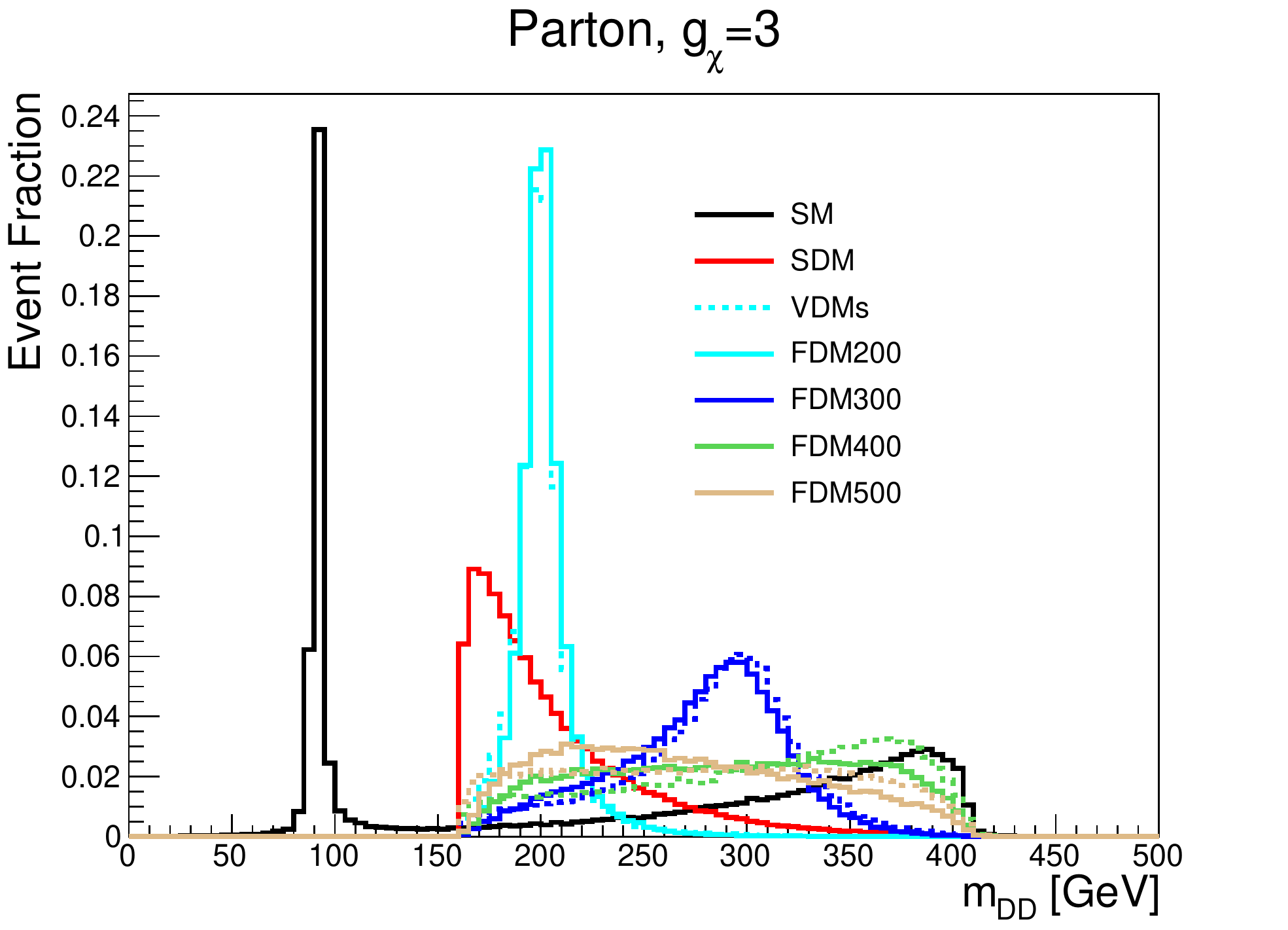} 
\includegraphics[width=0.4\textwidth]{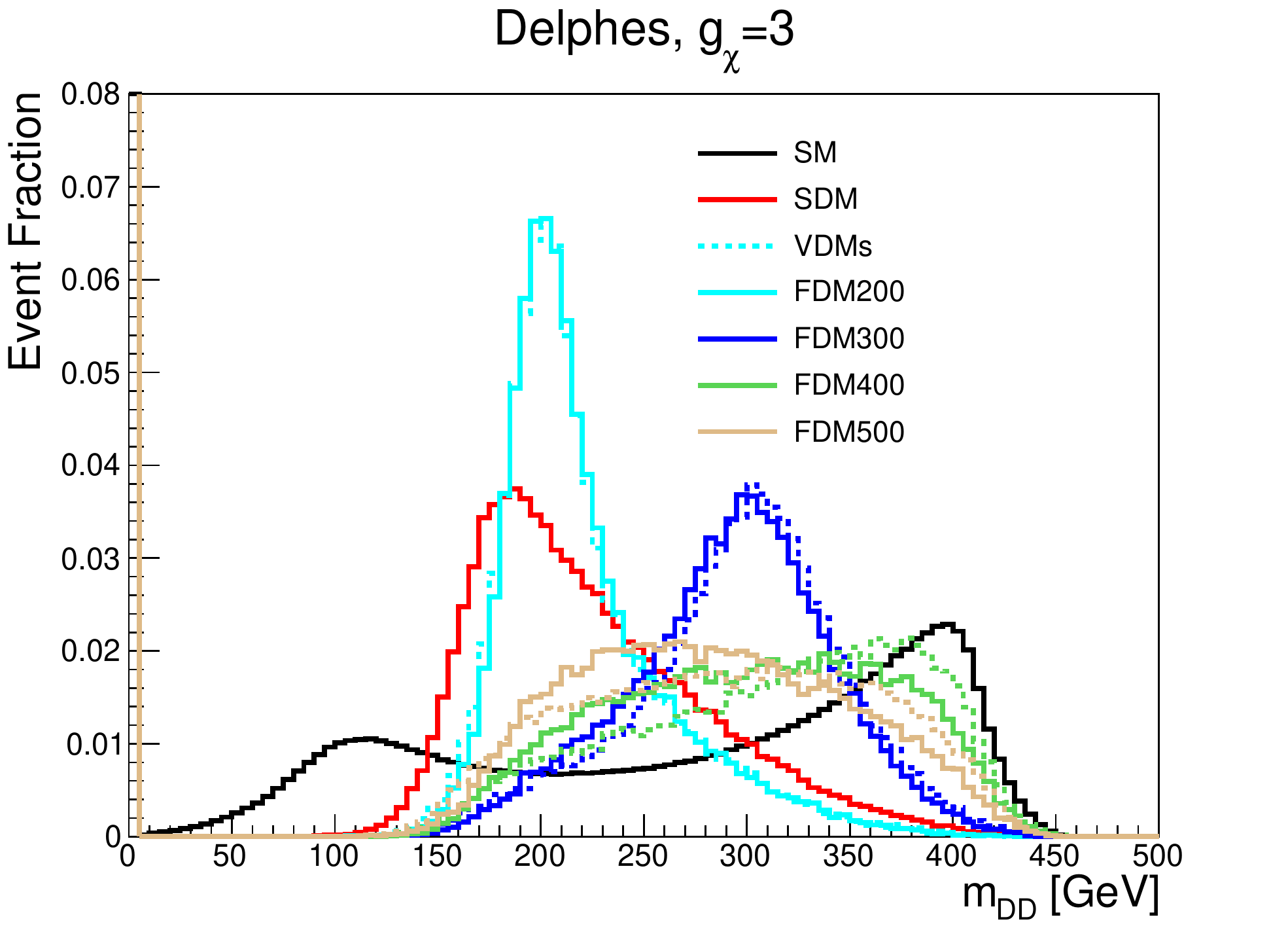}
\end{center}
\caption{\label{mdd} Invariant mass of DM (neutrino) pair for signal (background). Left panel shows parton level distributions. Right panel shows the detector level distributions. The meanings of lines are same as in Fig.~\ref{sigmas}.}
\end{figure}

For fixed $\sqrt{s}$, a powerful spin discriminator at the ILC is the invariant mass of the DM pair 
$m_{DD} \equiv \sqrt{t}$.  We plot the $m_{DD}$ distributions for signals with different DM spins as well 
the background both at parton level (left panel) and at detector level (right panel)  in Fig.~\ref{mdd}. 

At parton level, the $m_{DD}$ for SM background corresponds to the invariant mass of the neutrino pair in the final state, since they will mimic the missing energy from the DM pair at detector level. 
As we have discussed before, there is a large fraction of background events in which the neutrino pair is produced 
from $Z$ boson decays. Thus the $m_{DD}$ will show a sharp peak at $m_Z$ which is a SM background. 
The $m_{DD}$ is usually quite large for the VBF background process (first panel in Fig.~\ref{bgs}), which gives another broad peak at $m_{DD} \sim 400$ GeV.  In the SDM model, the DM with $m_S=80$ GeV is pair produced through
the off-shell SM Higgs mediation. The $m_{SS}$ will peak at $2 m_S$ and decrease as $1/m_{SS}^4$ with increasing $m_{SS}$. In FDM and VDM models, there is another resonant enhancement at $m_{DD} \sim m_{H_2}$ because of the existence of the additional scalar mediator, especially when the mass of $H_2$ is relatively light and decay width of the $H_2$ is small. This explains the clear peaks for FDM200 and VDM200. The peaks become 
much broader for $m_{H_2}=300$ GeV since the decay width of $H_2$ is large. As  the on-shell $H_2$ production 
is (almost) kinematically closed for $m_{H_2}=400/500$ GeV, the peaks no longer exist. 
The FDM and VDM also show distinguishable structures in the $m_{DD}$ distributions. 
When the second scalar mediator is light the VDM has more events in the small $m_{DD}$ region than 
the FDM while this becomes opposite when the second mediator is heavy. 

The features at parton level can be smeared out to some extent by the detector effects. First of all, the momenta of DMs/neutrinos are not observables. One can only calculate the $m_{DD}$ from Eq.~(\ref{eq:mdd}) by using the momentum of the $Z$ boson, which is identified as the vector sum of the momenta of two leading jets. In some cases, only one of two jets from the $Z$ boson decay is reconstructable at the detector ($p_T(j) >20$ GeV and $|\eta(j)| < 3.0$). These events will be dropped. The detector level distribution of the $m_{DD}$ is given in the right panel of Fig.~\ref{mdd}. We can see that the peaks are broadened and the edges get ambiguous. In particular, for the background process, the peak at $Z$ boson mass is almost disappearing and the distribution of $m_{DD}$ is 
quite flat, rending the discovery of signal processes difficult. The edges for signal distributions at $2 m_D$ 
and $\sqrt{s}-m_{Z}$ are less steep. 
Nevertheless, we are still able to observe distinguishable distributions between signal and background as well as between signals with different DM spins. These features can be used to search and characterize the signal as will be discussed in the following.

\subsection{Discovery prospect of FDM and spin discriminating power} 
\label{sec:fdmg3}

A signal has to be discovered with high significance before being characterized. 
In this section, based on the benchmark scenarios that we have set at the beginning of this section, we will study 
the discovery prospects of the FDM and discuss its spin discriminating power against SDM and VDM at the ILC 
with $\mathcal{L}=1000$ fb$^{-1}$ and  $\sqrt{s}=500$ GeV.  

In the event reconstruction, leptons are required to have $p_T(\ell)>10$ GeV, $|\eta(\ell)|<2.4$~\footnote{It would be more conventional to use variables of momentum $p$ and polar angle $\theta$ at electron positron collider, which is, however, not supported in Delphes yet. We will follow the notation as in Delphes ILD card with selections applied to $p_T$ and $\eta$ throughout the paper. It has to be noted that such a choice will not bring much differences into our final results because of the following reasons: (1) $\theta$ is simply given by $\theta = 2 \arctan (e^{-\eta})$; (2) the $p_T$ and $p$ are highly correlated, they have similar sensitivities in signal and background discrimination. } and be isolated which means  the scalar sum of transverse momenta of all particles with $p_T > 0.5$ GeV that lie within a cone of radius $R = 0.5$ around the $e(\mu)$ is less than 12\%(25\%) of the transverse momentum of the $e(\mu)$. 
Jets are reconstructed from particle flow objects from Delphes using the anti-kt jet clustering algorithm~\cite{Cacciari:2008gp} with a radius parameter $R = 0.5$. Only jet candidates with $p_T(j) > 20$ GeV and $|\eta| < 3.0$ are considered as signal jets in our analysis. The missing transverse momentum $p_T^{\text{miss}}$ is defined as the negative vector sum of the transverse momenta of all identified physics objects at the detector.  Candidate events should pass the preselection cuts: (1) no leptons in the final state; (2) exactly two jets in the final state; (3) $E^{\text{miss}}_T \equiv |p_T^{\text{miss}}| > 50$ GeV~\footnote{$E^{\text{miss}}_T$ is used instead of $E^{\text{miss}}$, because the imperfection of detecting particles that are close to the beam pipe may lead to artificial momentum imbalance along the longitudinal direction.}. 

The cross sections of the benchmark points in FDM model before and after the preselection are given in 
Table ~\ref{Fg3}, where we have taken into account the $Z$ boson hadronic decay branching ratio. 
It can be seen that the total cross section decreases quickly with increasing  the mediator mass. 
The preselection efficiency is relatively flat ($\sim 0.7-0.8$) and is smallest when $m_{H_2} =400$ GeV. This is because for each event, the DM pair recoil energy ($E^{\text{miss}}_T$) is in inverse proportion to the invariant mass of dark matter pair ($m_{DD}$). As can be seen clearly in the right panel of Fig.~\ref{mdd}, the distribution of $m_{DD}$ is hardest for FDM400, while it is decreasing with either larger or smaller $m_{H_2}$. 

On the other hand, the production cross sections of the SM background before and after the preselection are 
found to be 219 fb and 109.1 fb, respectively, which are typically more than two order of magnitude larger than 
that of our signals. Such small signals can be easily hidden in the background with relatively large uncertainty. 
One would rely on more refined cuts to improve the signal-to-background ratio as well as the signal significance.

\begin{figure}[htbp]
\begin{center}
\includegraphics[width=0.4\textwidth]{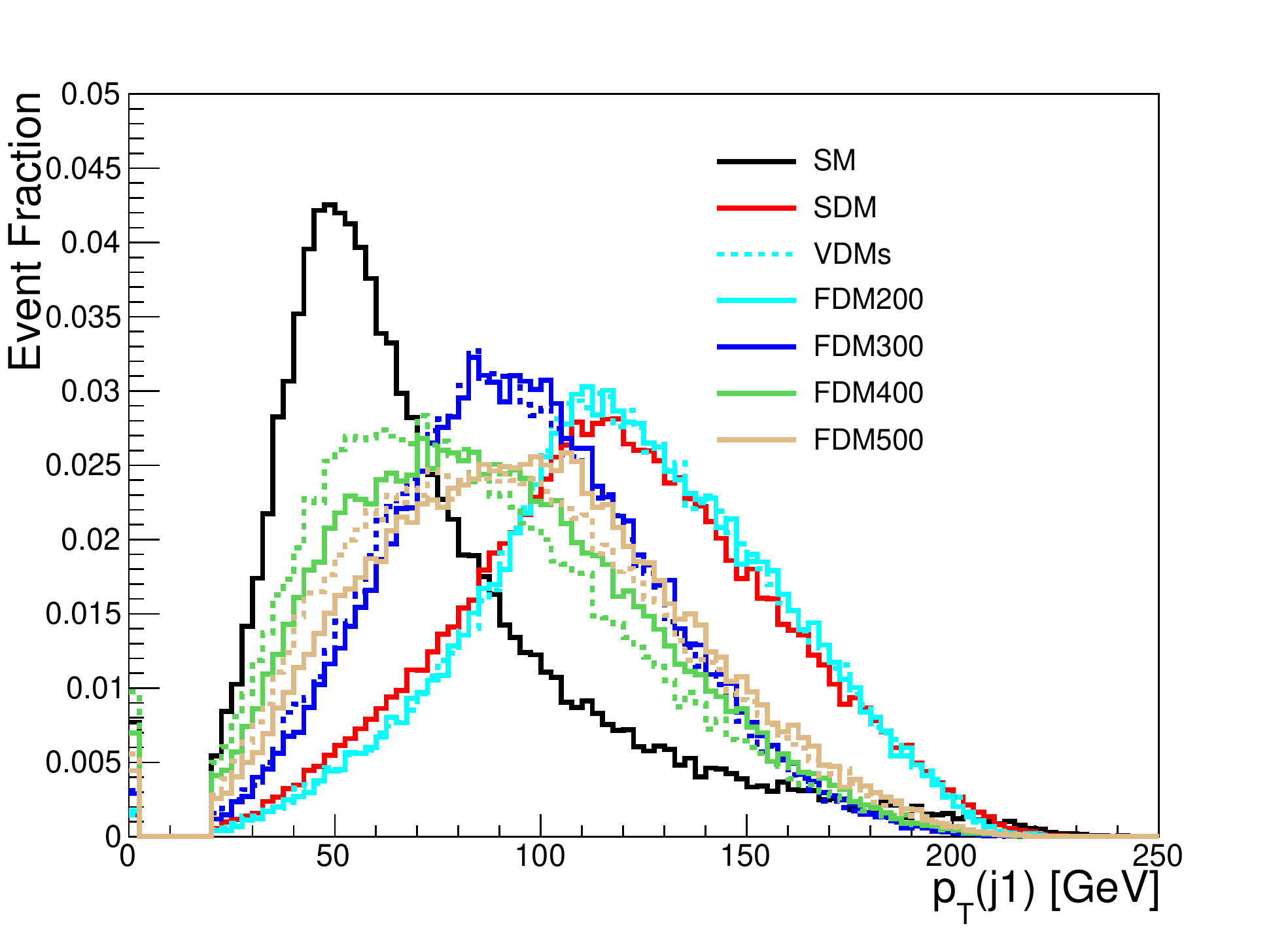} 
\includegraphics[width=0.4\textwidth]{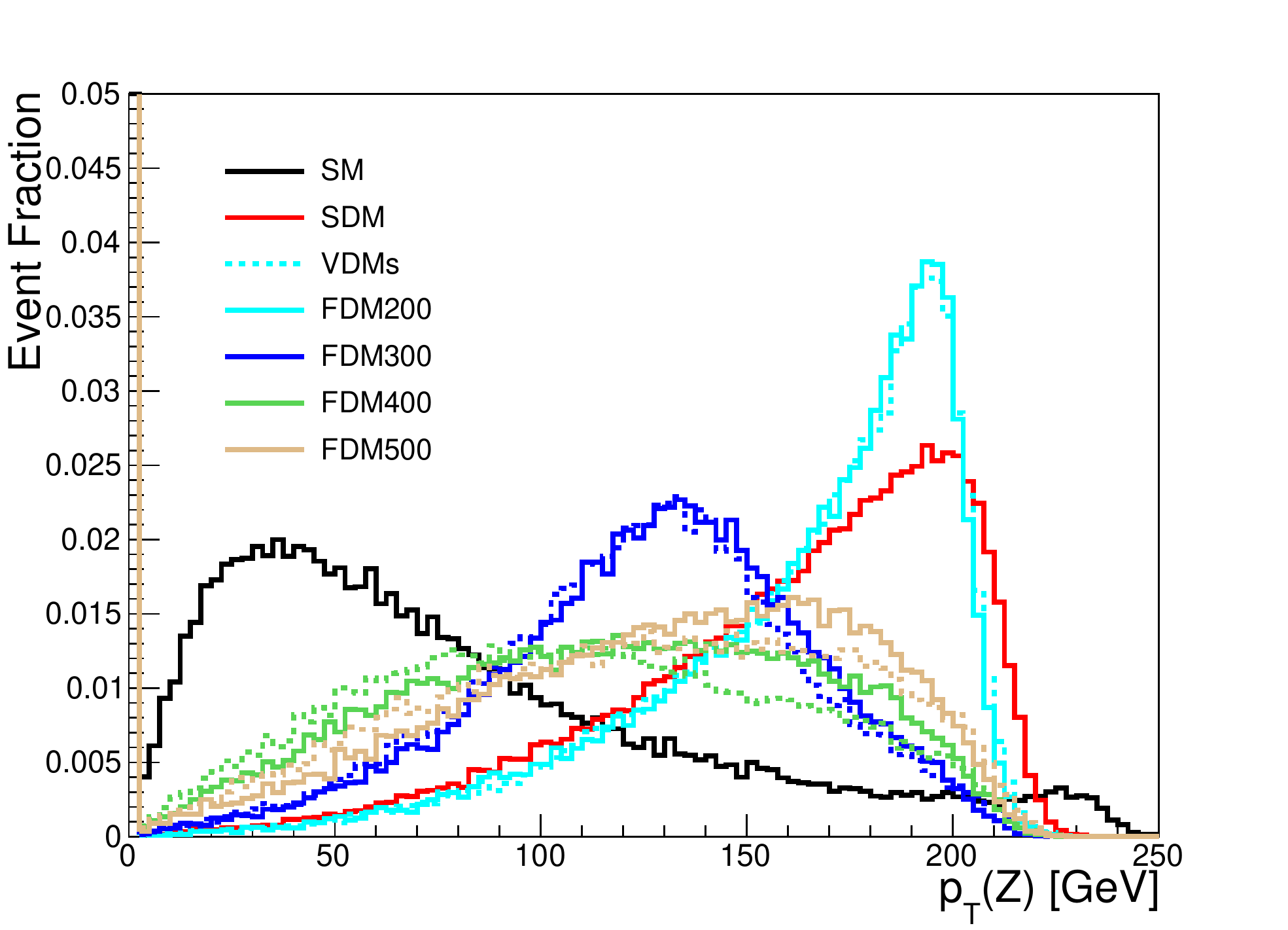} \\
\includegraphics[width=0.4\textwidth]{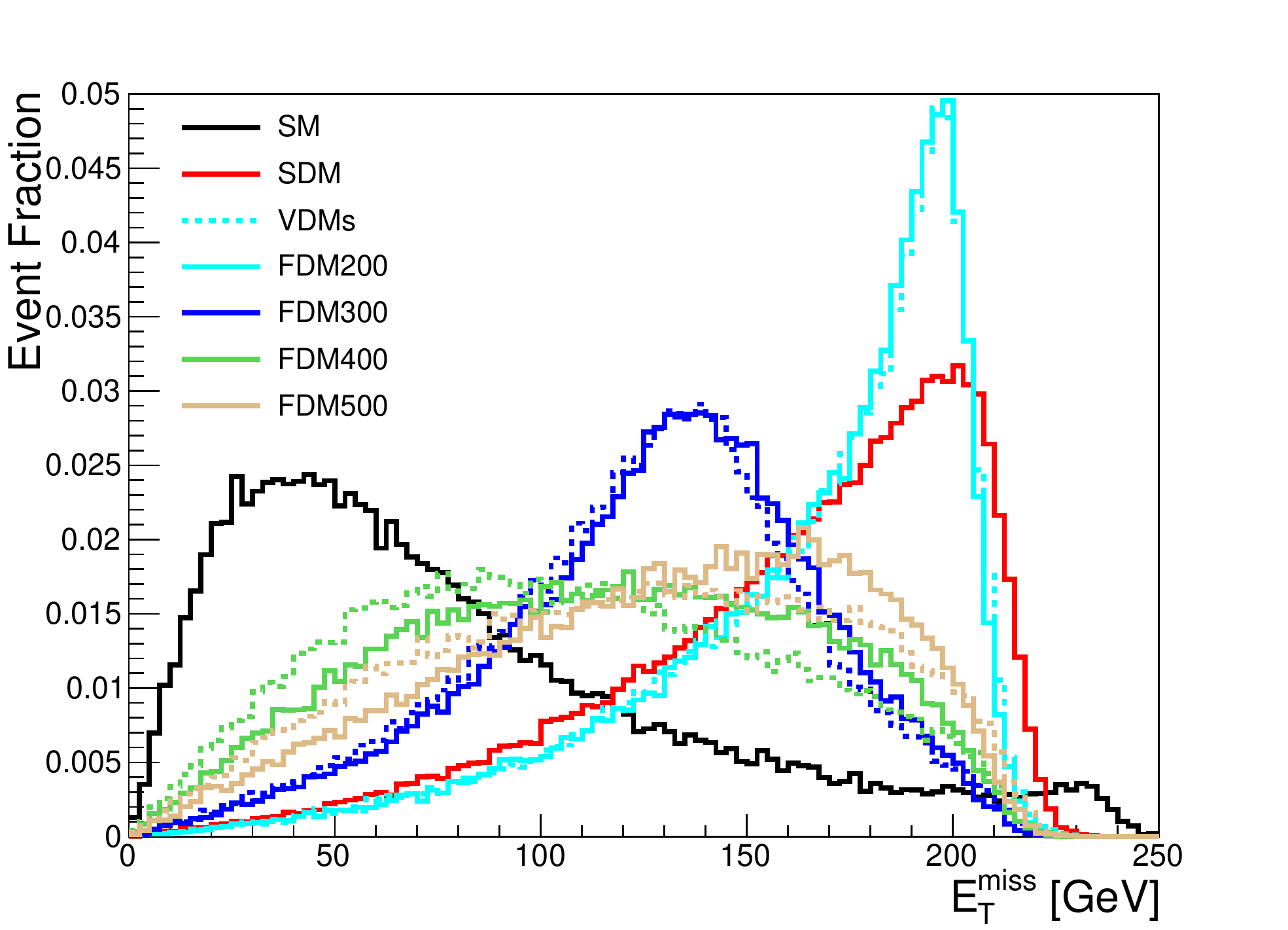} 
\includegraphics[width=0.4\textwidth]{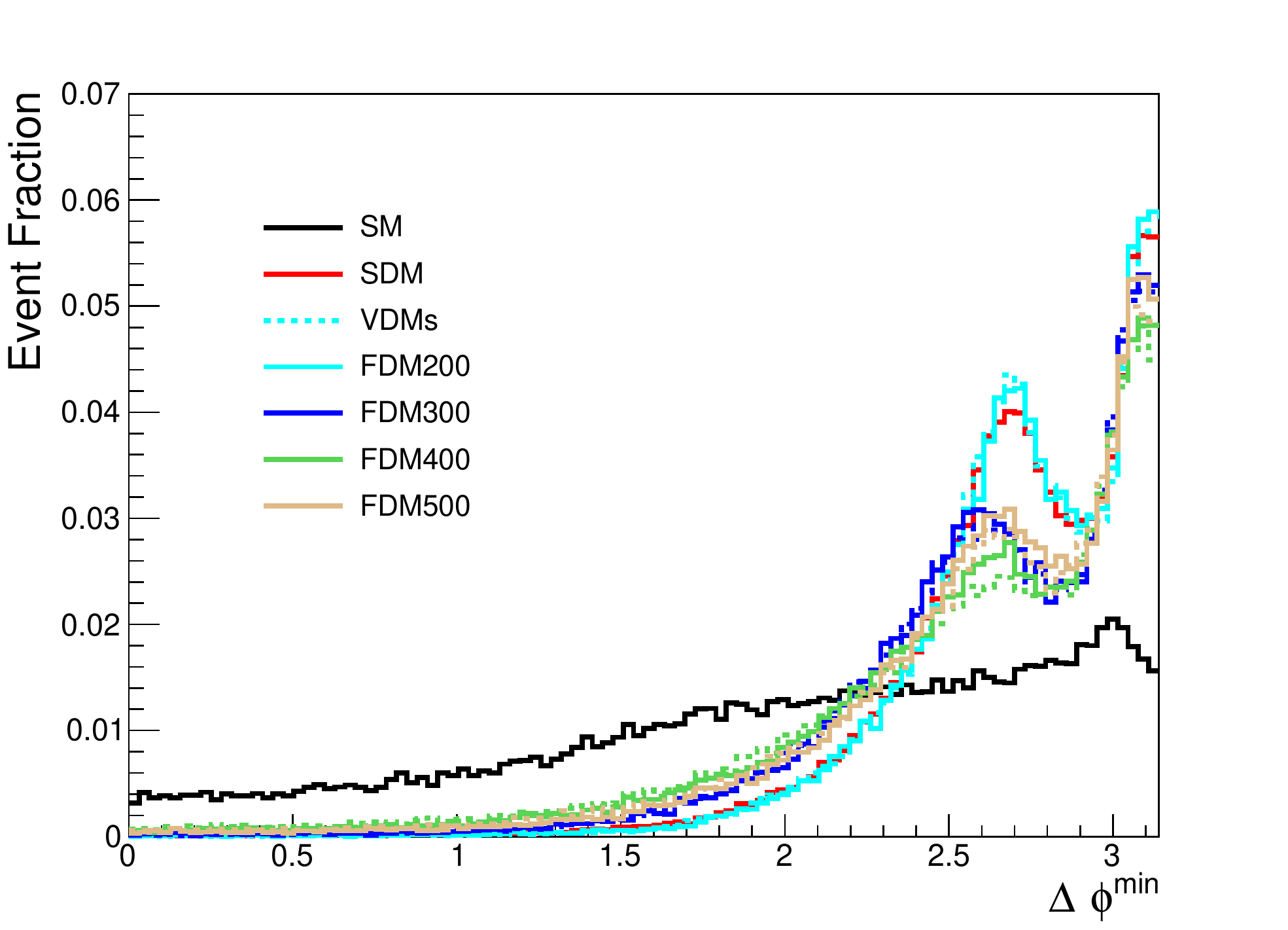}
\end{center}
\caption{\label{kins} Kinematic variables distributions after detector simulation. The meanings of lines are same as in Fig.~\ref{sigmas}.  }
\end{figure}

From the left panel of Fig.~\ref{mdd}, we know the $m_{DD}$ can play an important role in signal and background discrimination. Moreover, in signal processes, the DM pair is produced with recoiling against a Z boson which decays into two detectable jets. The two DM particles are flying along the similar direction. 
While in the background process, in particular the first and third diagrams in Fig.~\ref{bgs}, the momenta of two neutrinos are unlike to align with each other leading to a cancellation in missing transverse momentum. 
As a result, both the missing transverse energy ($E^{\text{miss}}_T$) and the transverse momentum of the 
$Z$ boson ($p_T(Z)$) and the leading jet ($p_T(j_1)$) get softened for the background, as being demonstrated 
in the Fig.~\ref{kins}. We note that the distributions of $E^{\text{miss}}_T$, $p_T(Z)$ and $p_T(j_1)$ are highly 
correlated: hardest for VDM200 and SDM; softest for VDM400 and SM background.

Another useful and less correlated discriminator is the azimuthal angle separation between the 
$p^{\text{miss}}_T$ and the momentum of the closer jets:
\begin{align}
\Delta \phi^{\min} = \min_{i=1,2} \Delta \phi \left(p^{\text{miss}}_T, p \left(j_i \right) \right)~.
\end{align}
In the signal process, the DM pair is flying around the opposite direction of an energetic $Z$ boson, which decays to two collinear jets. The $\Delta \phi^{\min}$ is distributed toward $\sim \pi$. As for background processes, 
where the $Z$ boson energy is much smaller, the  $\Delta \phi^{\min}$ distribution is flatter. 

We will adopt the BDT method~\cite{Roe:2004na,Hocker:2007ht} 
that takes into account all the above variables as well as the transverse momentum 
of the second leading jet ($p_T(j_2)$) and the invariant mass of jet pair ($m_{jj}$) in order to discriminate 
each signal benchmark point against the SM background. 
The BDT method uses a 100 tree ensemble that requires a minimum training events in each leaf node of 2.5\% and a maximum tree depth of three. For each benchmark point, it is trained on the half of the preselected signal and backgrounds events and is tested over the rest of the events. To avoid overtraining, the Kolmogorov-Smirnov test~\cite{Chakravarty:109749}  in the BDT training is required 
to be greater than 0.01. 

\begin{figure}[htbp]
\begin{center}
\includegraphics[width=0.4\textwidth]{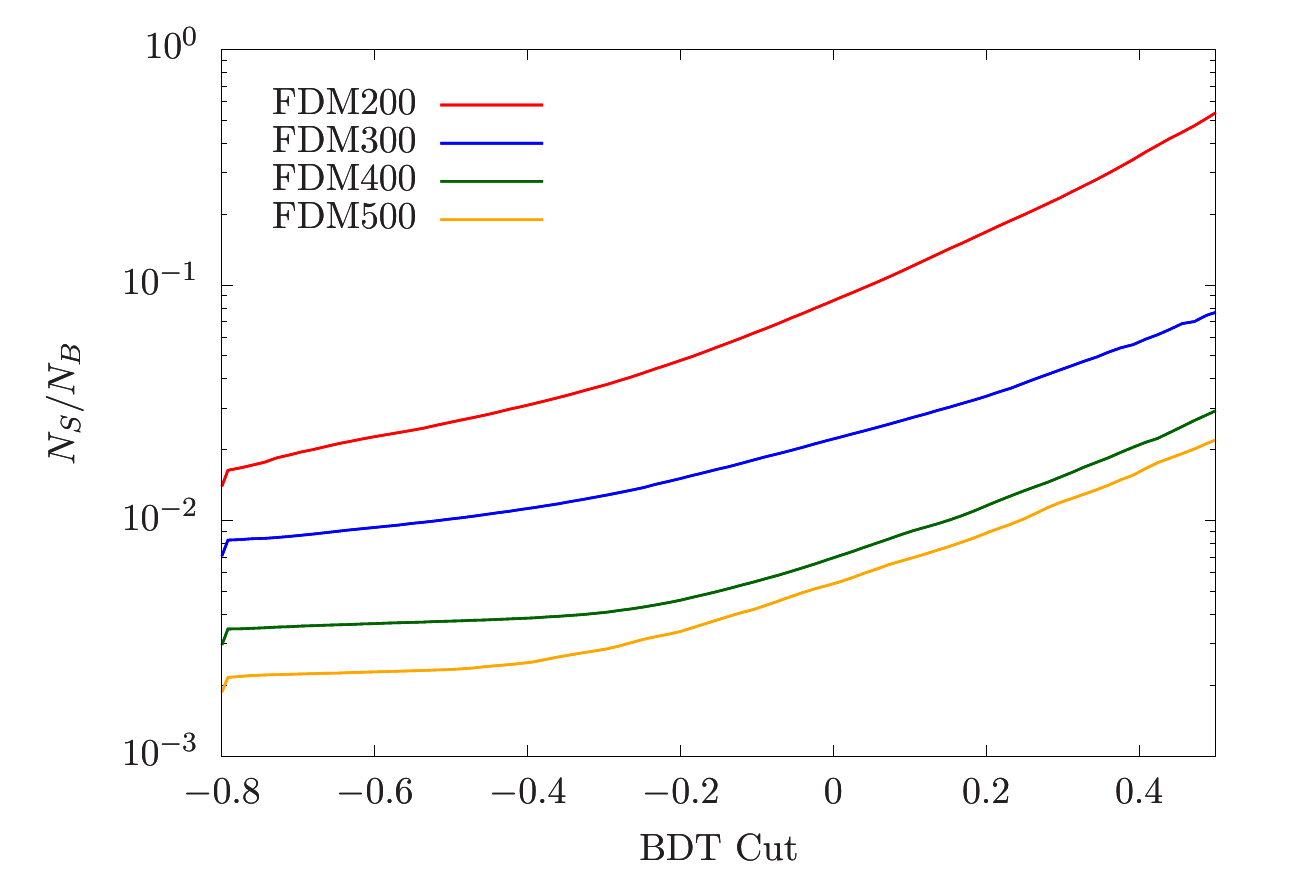} 
\includegraphics[width=0.4\textwidth]{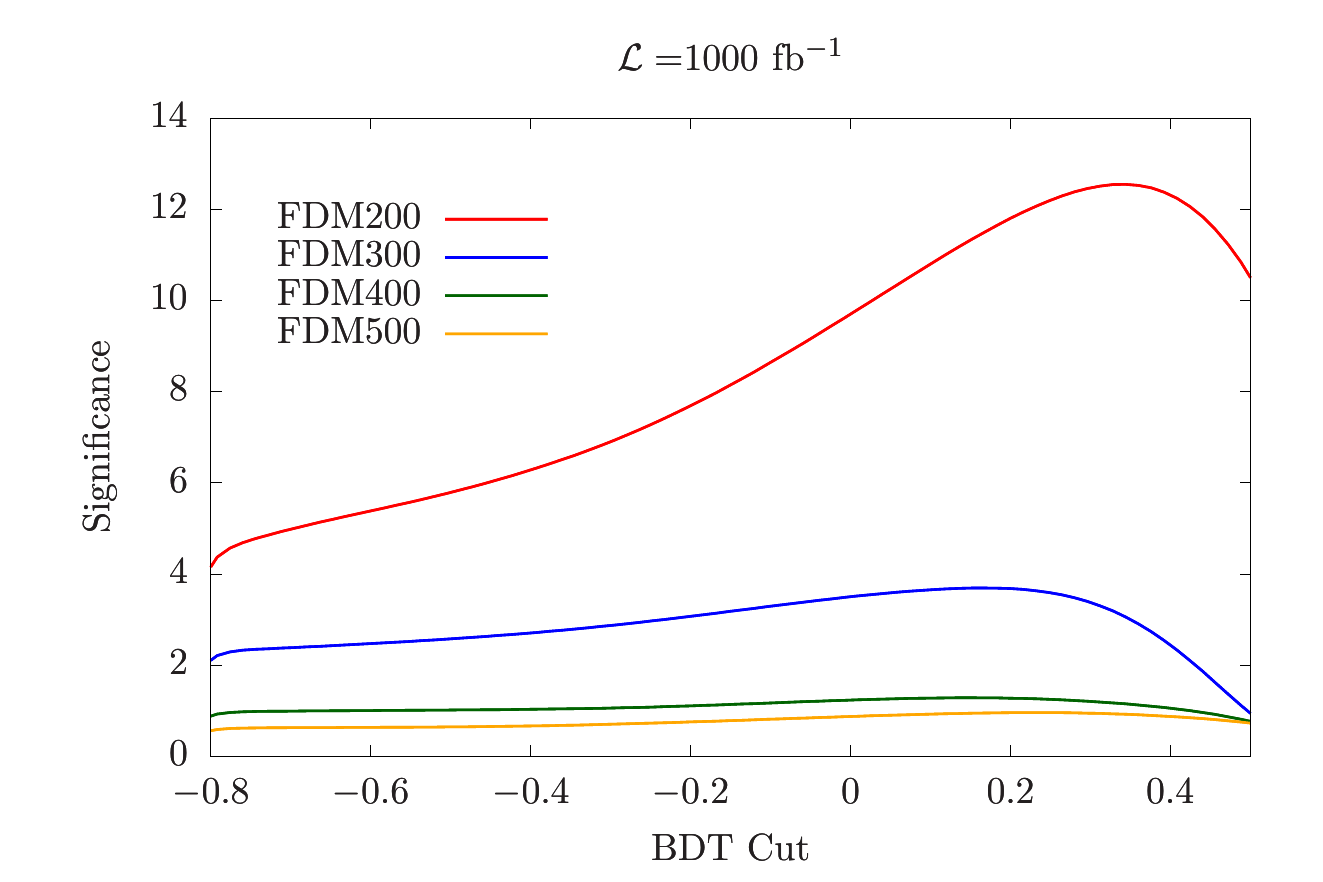}
\end{center}
\caption{\label{fig:bdtsig} Left: the signal-to-background ratio with varying BDT cut for FDM benchmark points. Right: the signal significance at the ILC with integrated luminosity of 1000 fb$^{-1}$. }
\end{figure}

After the BDT training, one can assign a BDT response to each event, which is usually larger for signal than for background. Distinguishable distributions of BDT response for signal and background can be obtained by taking into account a large number of events. Then, a cut on the BDT distribution can help to improve the signal purity. We plot the signal-to-background ratios ($N_S/N_B$) and the signal significances ($N_S/\sqrt{N_S+N_B}$) with varying cuts on the BDT distributions for FDM benchmark points in Fig.~\ref{fig:bdtsig}.  
We can see that the cuts on BDT can improve the $N_S/N_B$ by at least one order of magnitude, while improvements on the signal significance are only significant for benchmark points with relatively light mediator mass. 

The corresponding cut on BDT for each benchmark point in FDM model that maximizes the signal significance is given in the Table~\ref{Fg3}, where we also provide the numbers of signal and background events and the signal significance after the BDT cut.  We find that detections on the benchmark points of FDM200 and FDM300 can be made at 3-$\sigma$ level at the ILC with collision energy of $\sqrt{s}=500$ GeV and integrated luminosity of 1000 fb$^{-1}$. This would allow us to perform the spin discrimination for those two benchmark points.

\begin{table}[htb]
\begin{center}
\begin{tabular}{|c||c|c|c|c|c|} \hline
  & FDM200 & FDM300 & FDM400 & FDM500\\ \hline
$\sigma^0$ [fb] & 1.643 & 0.9214 & 0.4221 & 0.2526 \\  \hline
$\epsilon^{\text{pre}}$ & 0.796 & 0.717 & 0.655 & 0.698 \\ \hline 
BDT & 0.3615 & 0.2132 & 0.1929 & 0.2129 \\ \hline
$N_S$/1000 fb$^{-1}$ & 697.8 & 410.5 & 148 & 102 \\ \hline
$N_B$/1000 fb$^{-1}$ & 2248.5 & 11453.5 & 12736 & 10898  \\ \hline
$N_S/\sqrt{N_S+N_B}$ & 12.85 & 3.769 & 1.31 & 0.97 \\ \hline
\end{tabular}
\caption{\label{Fg3} The total production cross section ($\sigma^0$), cross section after pre-selection ($\epsilon^{\text{pre}}$), the chosen BDT cut (BDT), number of signal ($N_S$) and background ($N_B$) events after BDT cut and the signal significance ($N_S/\sqrt{N_S+N_B}$) at the ILC with $\sqrt{s}=500$ GeV and $\mathcal{L}=$1000 fb$^{-1}$ for benchmark points in FDM model.  }
\end{center}
\end{table}

The procedure of the spin discrimination can be described as the following. 
Firstly, events are simulated and production cross sections are calculated for benchmark points in SDM model (SDM200, SDM300) and in VDM model (VDM200, VDM300). The SDM200 (SDM300) denotes benchmark point in SDM model that has the same signal yields after the event selection as the FDM200 (FDM300) and the VDM200 (VDM300) denotes the benchmark point in VDM model that has the second mediator mass of 200 (300) GeV.  
Next, after the event reconstruction, the same preselection cuts as for FDM are applied. 
The cross sections as well as the preselection efficiencies for those benchmark points are provided in the 
Table~\ref{SVg3}. Note that the preselection efficiencies for SDM200 and SDM300 are the same, since 
the only free parameter $\lambda_{HS}$ in SDM model can not change the kinematic features of the final state.  
Then, we apply the BDT that has been trained on the benchmark point FDM200 (FDM300) to the corresponding benchmark point SDM200 (SDM300) and VDM200 (VDM300). 
Finally, we apply the BDT cuts as given in the fourth row of Table~\ref{Fg3} to the corresponding benchmark points 
in SDM and VDM model. The event numbers at $\mathcal{L}=1000$ fb$^{-1}$ for those benchmark points are given 
in the fourth row of Table~\ref{SVg3}. 

\begin{table}[htb]
\begin{center}
\begin{tabular}{|c||c|c|c|c|c|} \hline
  & SDM200 & SDM300 & VDM200 & VDM300\\ \hline
$\sigma^0$ [fb] & 2.56 & 1.17 & 1.734 & 0.8674 \\  \hline
$\epsilon^{\text{pre}}$ & 0.7875 & 0.7875 & 0.801 & 0.711 \\ \hline 
$N_S$/1000 fb$^{-1}$ & 697.8 & 410.5 & 726 & 363.5 \\ \hline
$\mathcal{S}$ & 2.54 & 4.53 & 0.59 & 0.44 \\ \hline
\end{tabular}
\caption{\label{SVg3} The total production cross section ($\sigma^0$), cross section after pre-selection ($\epsilon^{\text{pre}}$) and number of signal ($N_S$) at the ILC with $\sqrt{s}=500$ GeV and $\mathcal{L}=$1000 fb$^{-1}$ for benchmark points in SDM and VDM model. The last row gives the spin discriminating significances of FDM with $m_{H_2} =200/300$ GeV against corresponding SDM and VDM. }
\end{center}
\end{table}

\begin{figure}[htbp]
\begin{center}
\includegraphics[width=0.4\textwidth]{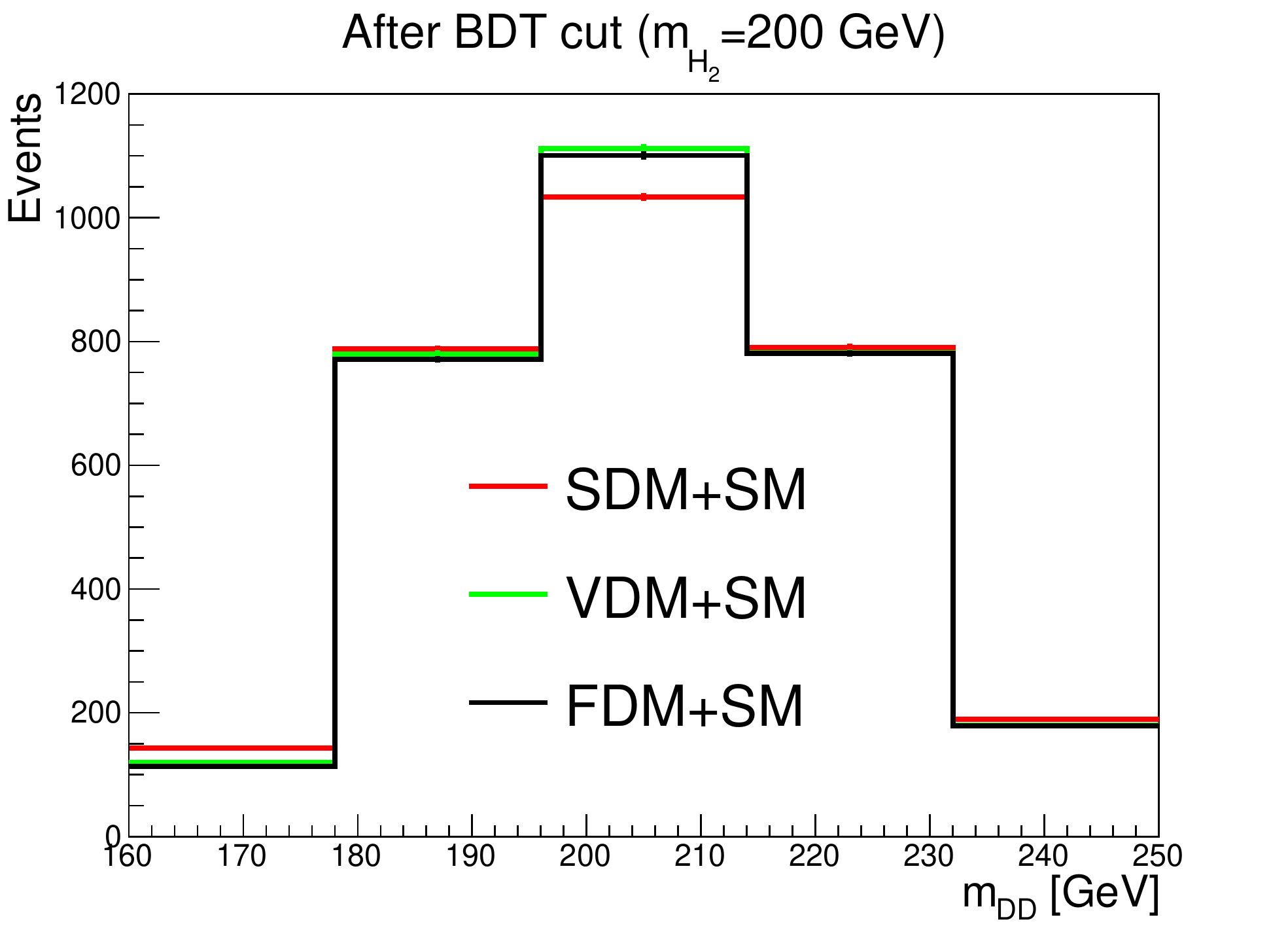} 
\includegraphics[width=0.4\textwidth]{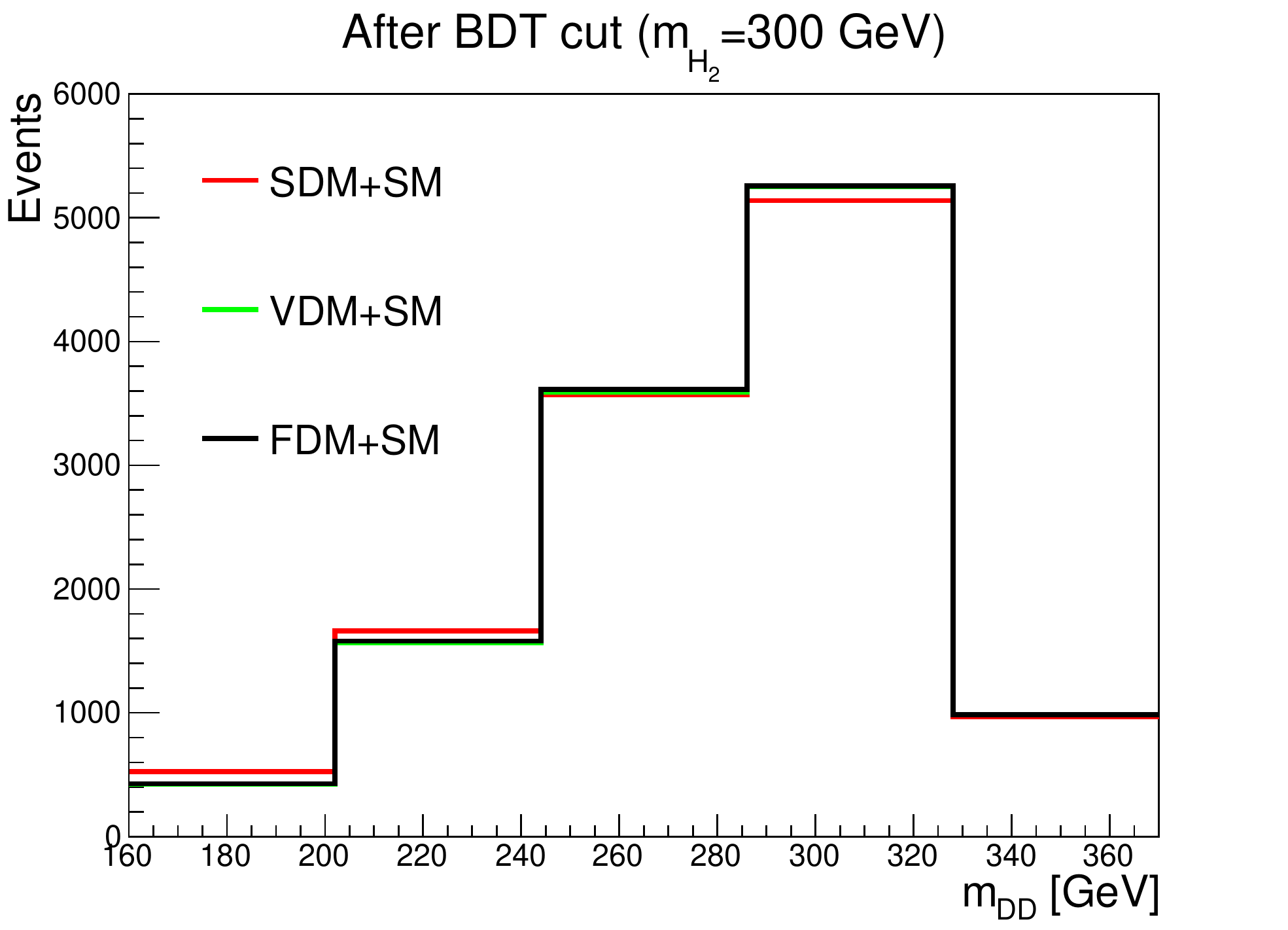}
\end{center}
\caption{\label{mddbdt} Distributions of $m_{DD}$ after the BDT cut for each signal plus background. Left: discriminating the spin of the benchmark point FDM200. Right: discriminating the spin of the benchmark point FDM300.   }
\end{figure}

The survived events are used to plot the distributions of $m_{DD}$ for different models. 
In Fig.~\ref{mddbdt}, we give the 5-bin distributions of $m_{DD}$ after applying the BDT cut for signals of different DM spin adding to the SM background. 
We can observe that the $m_{DD}$ distributions of benchmark points in FDM and SDM model have visible difference, while that of benchmark points in FDM and VDM are almost the same. 
To assess the degree of difference between the benchmark points in FDM and SDM, we construct the $\chi^2$ 
statistic 
\begin{align}
\delta \chi^2 = \sum_{i=1}^5 \left( \frac{N_i^{\text{FDM+SM}} - N_i^{\text{SDM+SM}}  }{\sqrt{N_i^{\text{FDM+SM}}}} \right)^2  \label{eq:chi2}
\end{align}
where $N_i^{\text{FDM+SM}}$ ($N_i^{\text{SDM+SM}}$) is the number of FDM (SDM) signal plus background events in the $i$-th bin and the $i$ runs over five bins of the histograms in Fig.~\ref{mddbdt}. 
The $\delta \chi^2$ value is compared to the $\chi^2$ distribution with 4 degrees of freedom to calculate the $p$-value, which can be further transformed to the significance level ($\mathcal{S}$) from a Gaussian distribution. The $\mathcal{S}$ for each benchmark point in SDM model is given in the fifth row of Table~\ref{SVg3}. Both benchmark points in SDM model can be distinguished from the benchmark points in FDM at significance level of more than 2-$\sigma$. 
We note that the number of events after the BDT cut contains not only the information of normalization of the 
$m_{DD}$ distribution but also the information of its shape, since the BDT selection used the $m_{DD}$ distribution.
Therefore, for discriminating FDM and VDM, the significance level will be simply estimated by $\mathcal{S}= \left|N_S^{\text{FDM}} - N_S^{\text{VDM}} \right|/\sqrt{N_B}$, with $N_S^{\text{FDM}}$ ($N_S^{\text{VDM}}$) is the number of FDM (VDM) signal events as given in Table~\ref{Fg3} (Table~\ref{SVg3}), $N_B$ is the number of background events after applying BDT cut. We find both benchmark points in VDM model can only be distinguished from the benchmark points in FDM with significance level below 1-$\sigma$.

\subsection{DM properties of benchmark points} 

In this subsection, we will briefly discuss the DM relic density~\cite{Ade:2015xua} and DM direct detection bound~\cite{Akerib:2016vxi} for our benchmark points~\footnote{Global analysis of fermion and vector DM with Higgs portal will be reported elsewhere~\cite{workprogress}.}. These values are calculated numerically by micrOMEGAs~\cite{Barducci:2016pcb} with the CalcHep/CompHEP~\cite{Christensen:2009jx} model files that are written by FeynRules~\cite{Alloul:2013bka,Christensen:2008py}. 
For all benchmark points, the DMs are dominantly annihilatting into $W W^{*}$ through scalar mediator(s) where $W^*$ is the off-shell $W$ boson. Due to the relatively large couplings between the mediator and DMs being chosen, the relic abundances of our DM particles are always below the measurement ($\Omega h_0^2 = 0.1198$) as can be seen in Tab.~\ref{dmbps}, rendering our DM particle only as a component of a full DM sector. Among DM spins, the fermion DM has suppressed s-wave annihilation, thus largest relic density. 

In comparison between the DM-proton scattering cross section in our model and the LUX constraint, the cross section ($\sigma_p^{\text{SI}}$) calculated in micrOMEGAs should be rescaled by a factor of $\frac{\Omega h^2}{0.1198}$ with $\Omega h^2$ being the calculated relic density of each benchmark point. According to Ref.~\cite{Akerib:2016vxi}, the current LUX measurement has excluded $\sigma_p^{\text{SI}} \cdot \frac{\Omega h^2}{0.1198} > 1.4 \times 10^{-10}$ pb for $m_{\text{DM}} = 80$ GeV, which means all of our benchmark points should have been excluded already. 
However, the direct detection limits rely on assumptions about the local dark matter density and velocity distributions, which are expected to vary from the standard assumptions used in the experimental results~\cite{Kuhlen:2009vh,Lisanti:2010qx,Mao:2013nda,Kuhlen:2013tra}. 
Moreover, if there is indeed a DM sector, our DM particle can either decay or annihilated into other dark particles, so that the direct detection can be evaded. It should be noted that those modifications will not lead to any effects in the collider phenomenology of DM searches. 

\begin{table}[htb]
\begin{center}
\begin{tabular}{|c|c||c|c|c|c|c|} \hline
\multicolumn{2}{|c||}{$m_{H_2}$ [GeV]}  & 200 & 300 & 400 & 500\\ \hline \hline
\multirow{ 2}{*}{FDM} & $\Omega h^2$ & $7.18 \times 10^{-3}$  & $1.18 \times 10^{-2}$ & $1.28 \times 10^{-2}$ &  $1.33 \times 10^{-2}$  \\  \cline{2-6}
& $\sigma^{\text{SI}}_p \cdot \frac{\Omega h^2}{0.1198}$ [pb] & $2.28 \times 10^{-9}$ & $1.13 \times 10^{-8}$ & $1.61 \times 10^{-8}$ & $1.87 \times 10^{-8}$   \\  \hline \hline
\multirow{ 2}{*}{VDM} & $\Omega h^2$ &  $4.78 \times 10^{-4}$ &  $1.60 \times 10^{-3}$ &  $3.05 \times 10^{-3}$ &  $4.88 \times 10^{-3}$ \\  \cline{2-6}
& $\sigma^{\text{SI}}_p \cdot \frac{\Omega h^2}{0.1198}$ [pb] & $8.44 \times 10^{-10}$ & $3.93 \times 10^{-9}$ & $5.32 \times 10^{-9}$ & $5.97 \times 10^{-9}$    \\  \hline \hline
\multirow{ 2}{*}{SDM} & $\Omega h^2$ & $2.83 \times 10^{-5}$ & $4.95 \times 10^{-5}$ & $1.04 \times 10^{-4}$ & $1.72 \times 10^{-4}$ \\  \cline{2-6}
& $\sigma^{\text{SI}}_p \cdot \frac{\Omega h^2}{0.1198}$ [pb] & $3.02 \times 10^{-9}$ & $2.94 \times 10^{-9}$ & $2.85 \times 10^{-9}$ & $2.78 \times 10^{-9}$   \\  \hline
\end{tabular}
\caption{\label{dmbps}  Relic densities and direct detection rates of benchmark points. }
\end{center}
\end{table}

\section{The leptonic channel}
\label{sec:g3lep}

As we have seen in Fig.~\ref{mdd}, the hadronic channel is suffering from the large uncertainty in jet momentum measurement, leading to smearing effects in the $m_{DD}$ distributions. 
On the other hand, much better lepton ($e/\mu$) momentum resolution of the leptonic channel may help to improve the discovery sensitivity as well as the spin discriminating power. 

However, the main drawbacks of the leptonic channel are its small production cross section and relatively large SM background. The $Z$ boson in the leptonic channel is required to decay into electron or muon pair, the decay branching ratio of which is around one order of magnitude below that of hadronic mode: Br$(Z\to \ell^+\ell^-)$ = 6.7\% with $\ell= e, \mu$, Br$(Z\to qq)$ = 69.9\% with $q=u,d,c,s,b$. Moreover, aside from the background processes listed in Fig.~\ref{bgs} with $j$ being replaced by $\ell$, there are new SM backgrounds such as the single $W$ and $W$ boson pair productions where the $W$ bosons are decaying leptonically. The total production cross section of the SM process $e^+ e^- \to \ell \ell \nu\nu$ is 505 fb at the $\sqrt{s}=500$ GeV ILC, which we find is dominated by the contributions from processes with $W$ boson in the final state.

\begin{figure}[htbp]
\begin{center}
\includegraphics[width=0.4\textwidth]{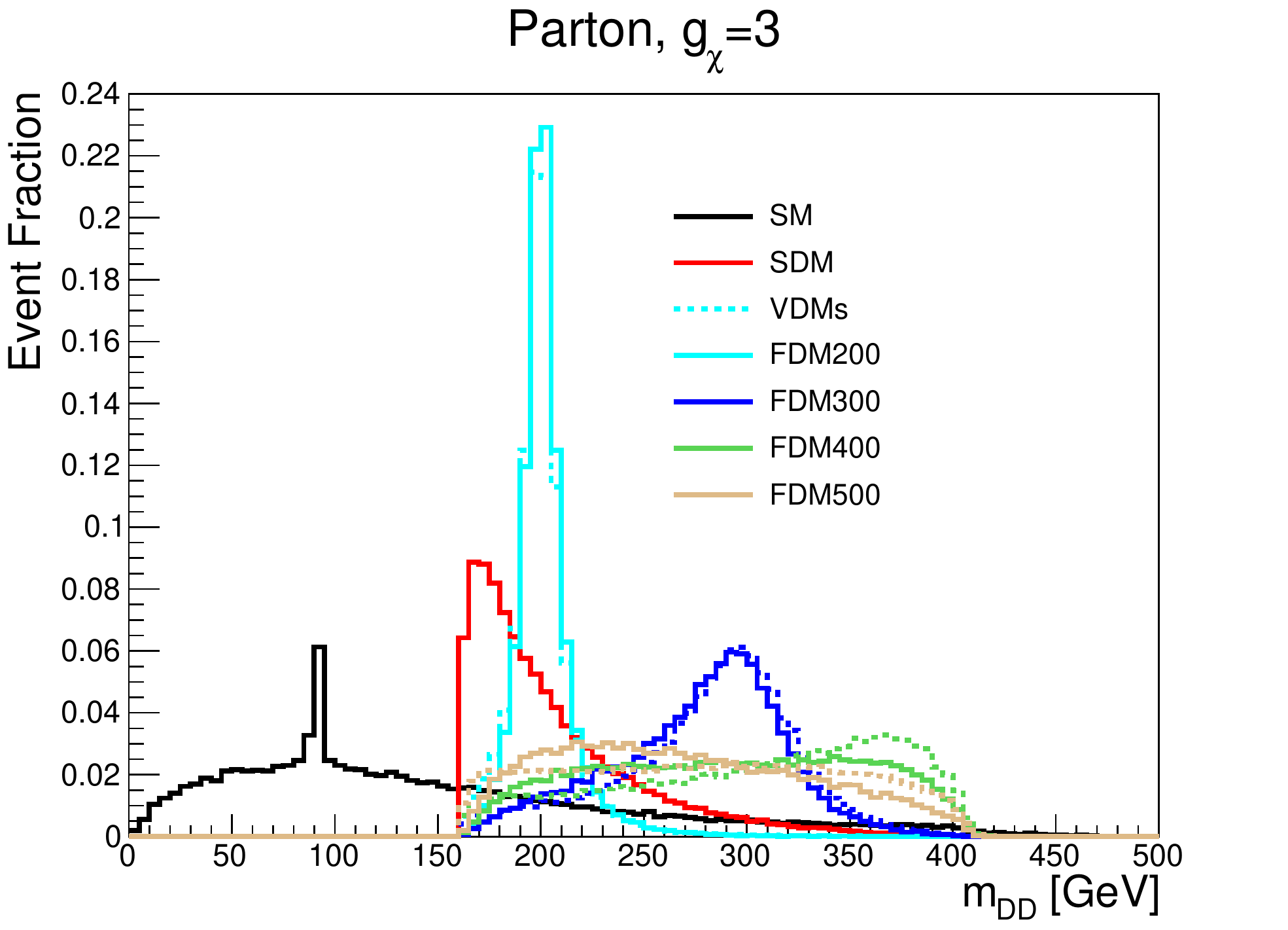} 
\includegraphics[width=0.4\textwidth]{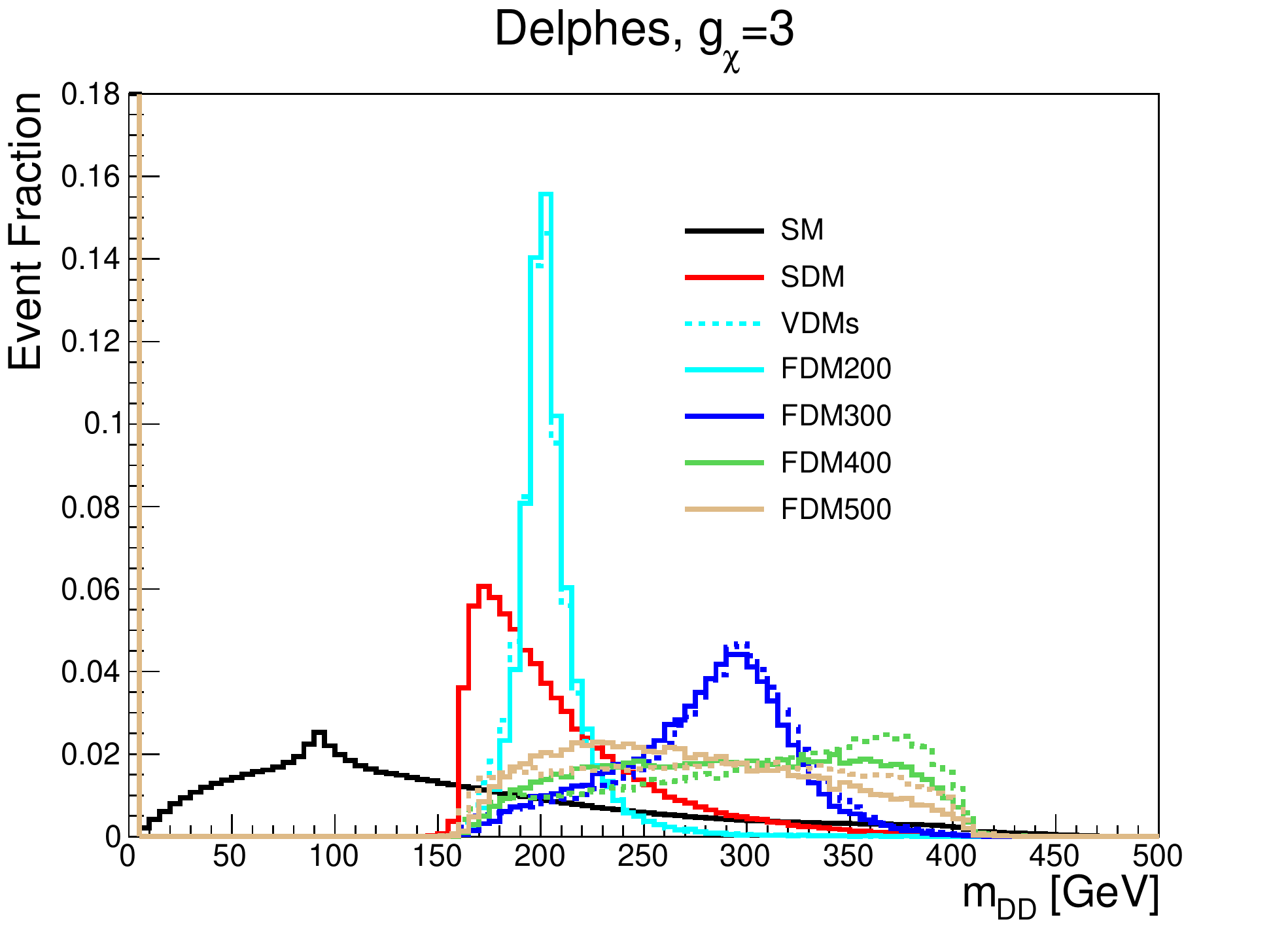}
\end{center}
\caption{\label{partong3ll} Invariant mass of DM (neutrino) pair for signal (background) in the leptonic channel. Left panel shows parton level distributions. Right panel shows the detector level distributions. The meanings of lines are same as in Fig.~\ref{sigmas}.}
\end{figure}

In Fig.~\ref{partong3ll}, we plot the $m_{DD}$ distribution for the leptonic channels of signals and background 
at parton level (left panel) and detector level (right panel). We can find that the shapes of $m_{DD}$ distributions 
are largely unaltered after taking into account the detector effects, i.e. peaks are sharp and edges are steep 
even at the detector level. 
Comparing to the Fig.~\ref{mdd}, the main features of signal distributions are kept the same as that in the hadronic channel, since the two channels only differ in the $Z$ boson decay final state. As for background, the $Z$ peak in the leptonic channel is less notable because the processes with $W$ in the final state are dominating. 
We note that in some events, only one of the two leptons in the final state is reconstructable at the detector ($p_T(\ell)>10$ GeV and $|\eta(\ell)|<2.5$). Those events are corresponding to those with $m_{DD}=0$ GeV in the right panel of Fig.~\ref{partong3ll}. 

Events for the leptonic channel are reconstructed with the same method as adopted for the hadronic channel. The candidate signal events are selected with the following preselection cuts: (1) exactly two opposite sign same flavor leptons in the final state; (2) no jet in the final state; (3) $E^{\text{miss}}_T > 50$ GeV; (4) two leptons invariant mass around the $Z$ pole $m_{\ell \ell} \in [75,105]$ GeV; (5) DM pair invariant mass above twice of the DM mass $m_{DD} > 160$ GeV. 
The total cross sections of the leptonic channels of benchmark points in the FDM model and their preselection efficiencies are given in Table~\ref{Fg3ll}. We also find the corresponding preselection efficiency of background 
is $\sim 0.029$ which is much smaller than that of signal. Nevertheless, after the preselection, the production rates 
of our signals are still around 2-3 order of magnitude smaller below that of the background.

\begin{table}[htb]
\begin{center}
\begin{tabular}{|c||c|c|c|c|c|} \hline
Leptonic channel  & FDM200 & FDM300 & FDM400 & FDM500\\ \hline
$\sigma^0$ [fb] & 0.2101 & 0.1181 & 0.0541 & 0.0323 \\  \hline
$\epsilon^{\text{pre}}$ & 0.722 &  0.703 & 0.652 & 0.677  \\ \hline 
BDT & 0.3775 & 0.25 & 0.26 & 0.335  \\ \hline
$N_S$/1000 fb$^{-1}$ & 85 & 47 & 16 & 9.72 \\ \hline
$N_B$/1000 fb$^{-1}$ &  151 & 1395 & 1376 & 830 \\ \hline
$N_S/\sqrt{N_S+N_B}$ & 5.5 & 1.24 & 0.43 & 0.34 \\ \hline
\end{tabular}
\caption{\label{Fg3ll} The meaning of each row is the same as in Table~\ref{Fg3}, with the leptonic channel instead. }
\end{center}
\end{table}

To increase the signal significance, we follow the similar strategy as in the hadronic channel, i.e. adopting the BDT method. The discriminating variables that are used in the leptonic channels are 
\begin{align}
p_T\left(\ell_1 \right), ~p_T \left(\ell_2 \right), ~E^{\text{miss}}_T, ~m_{\ell\ell}, ~ m_{DD}, ~p_T\left(Z \right), ~\Delta r \left( \ell,\ell \right), ~\Delta \phi^{\min},
\end{align}
where the $\Delta r\left(\ell,\ell \right) \equiv \sqrt{\left(\Delta \eta \left(\ell,\ell \right) \right)^2+ \left( \Delta \phi \left(\ell,\ell \right)  \right)^2}$ is the angular distance between two leptons and $\Delta \phi^{\min} \equiv \min_{i=1,2} \Delta \phi \left(p^{\text{miss}}_T, p\left(\ell_i \right) \right)$ is the azimuthal angular separation between the missing transverse momentum and the closer lepton. 

After training the BDT on each benchmark point in the FDM model, we can obtain the distributions of BDT response for signal and background. The cut on the BDT distributions is chosen such that the signal significance ($N_S/\sqrt{N_S+N_B}$) of each benchmark points is maximized. The corresponding BDT cuts, the number of signal and background events as well as the signal significance after BDT cuts are given in Table~\ref{Fg3ll}. Only the FDM200 
is discoverable at the ILC with $\sqrt{s}=500$ GeV and $\mathcal{L}=1000$ fb$^{-1}$. For all benchmark points, the signal significances of the leptonic channel are 2-3 times smaller than those of the hadronic channel. 

We can also discuss the spin discriminating of the FDM against SDM and VDM for the benchmark point FDM200. 
The production cross sections and preselection efficiencies of benchmark points SDM200 and VDM200 are given in the second and third row of Table~\ref{SVg3ll}. 
As in the hadronic channel, the significance levels ($\mathcal{S}$) of spin discriminations between FDM and SDM 
and between FDM and VDM are calculated with two different methods. The results are given in the fifth row of Table \ref{SVg3ll}. The FDM200 can be distinguished from SDM200 at significance level of around 2-$\sigma$, while it is impossible to be distinguished from VDM200.  
We can conclude that the hadronic channel provides better sensitivities in both signal discovery and spin discrimination than the leptonic channel.

\begin{table}[htb]
\begin{center}
\begin{tabular}{|c||c|c|c|c|c|} \hline
 Leptonic channel & SDM200  & VDM200\\ \hline
$\sigma^0$ [fb] & 0.504 & 0.2217  \\  \hline
$\epsilon^{\text{pre}}$ & 0.716 & 0.726 \\ \hline 
$N_S$/1000 fb$^{-1}$ & 85.0 & 88.1  \\ \hline
$\mathcal{S}$ & 2.31 & 0.25 \\ \hline
\end{tabular}
\caption{\label{SVg3ll} The meaning of each row is the same as in Table~\ref{SVg3}, with the leptonic channel instead.}
\end{center}
\end{table}

\section{Varying the coupling in the hadronic channel}
\label{sec:g110had}

So far, we have studied the benchmark points with $g_\chi=3$ in the FDM model. In this section, we will survey the discovery and spin characterizing prospects of benchmark points with $g_\chi=1$ and $g_\chi=10$ in the FDM model, while keep $\sin\alpha$ and $m_\chi$ unchanged. 
For each $g_\chi$, four different choices of $m_{H_2}$ = (200, 300, 400, 500) GeV will be considered. As have been done for the $g_\chi=3$ case, the corresponding benchmark points in VDM model are chosen such that the decay widths of $H_2$ are kept the same as the ones in the FDM model. We note that the branching ratio of $H_2 \to H_1 H_1$ is assumed to be negligible in calculating the decay width of a $H_2$. Benchmark points in the SDM model are chosen with the criterion that the signal yields after the event selection for signal process is the same with that of benchmark points in the FDM model by tuning the free parameter $\lambda_{HS}$.

\begin{figure}[htbp]
\begin{center}
\includegraphics[width=0.4\textwidth]{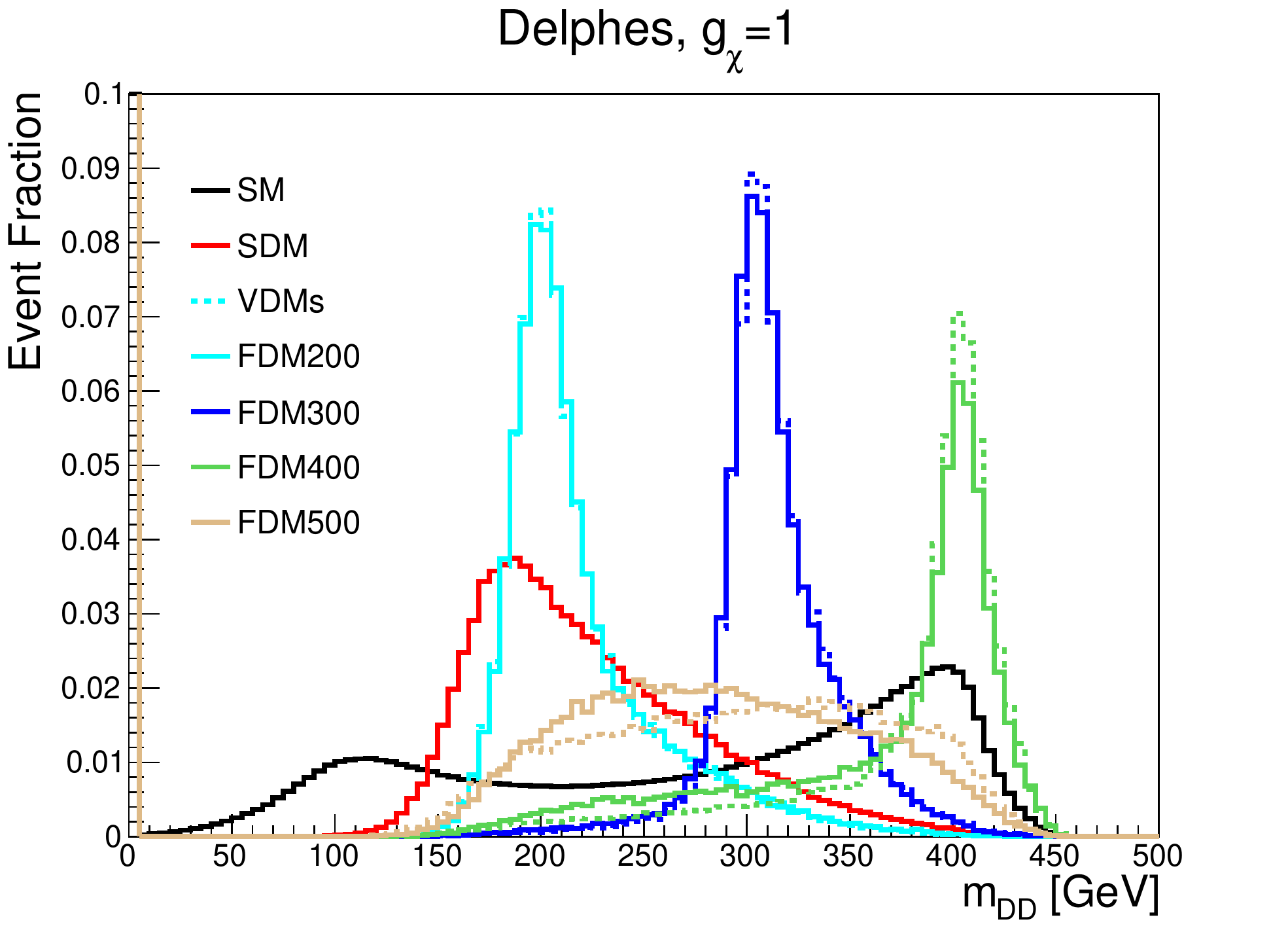} 
\includegraphics[width=0.4\textwidth]{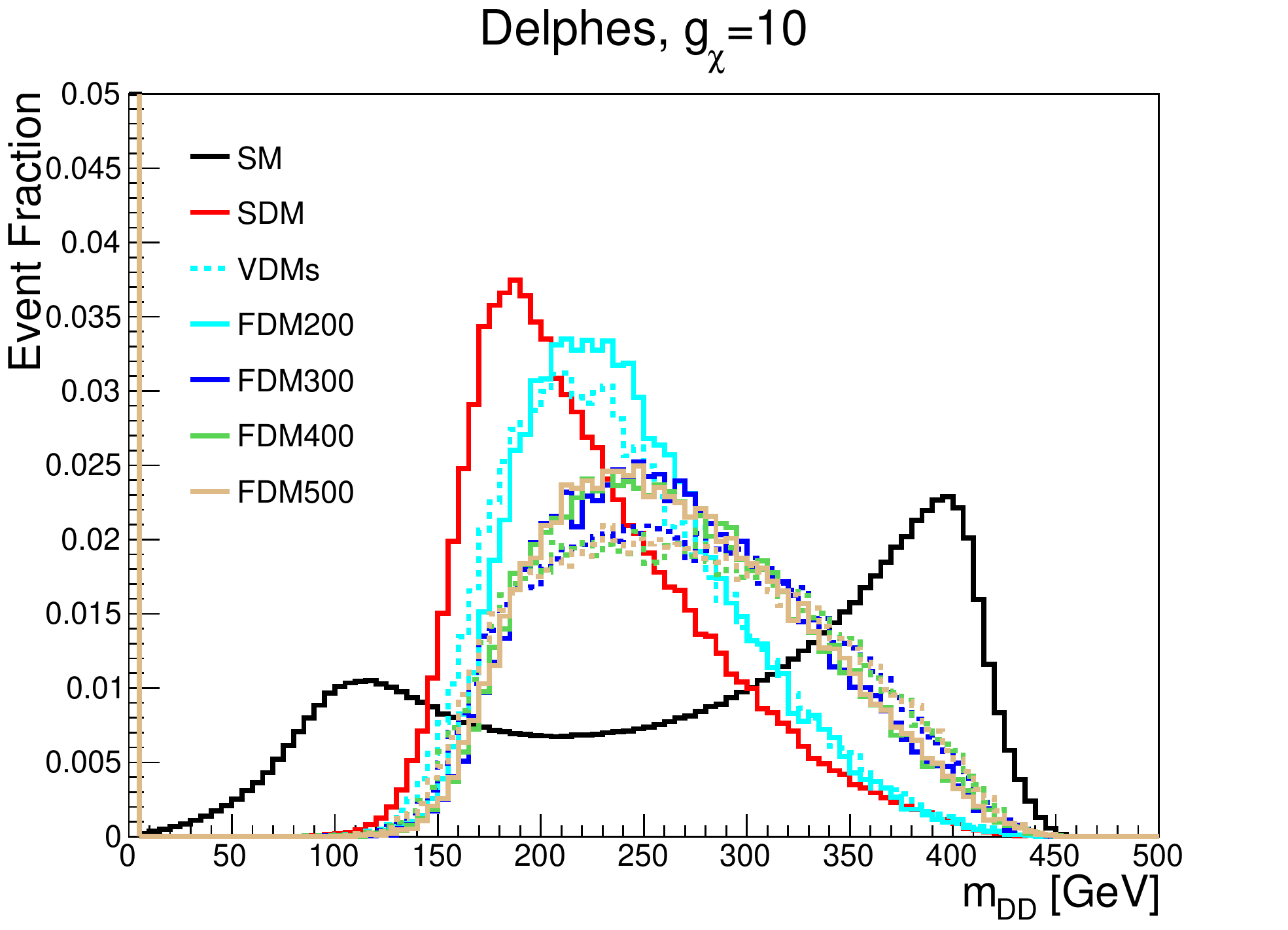}
\end{center}
\caption{\label{mddg110} Invariant mass of DM (neutrino) pair for signal (background) corresponding to two different choices of $g_\chi$ in FDM model. Left panel: $g_\chi=1$. Right panel: $g_\chi=10$. 
The meanings of lines are same as in Fig.~\ref{sigmas}. }
\end{figure}

The most important effect of changing the $g_\chi$ is that the total decay widths of the $H_2$ become different in the FDM and VDM models. As shown in Fig.~\ref{mddg110}, for FDM and VDM, the peaks in the $m_{DD}$ distribution are quite sharp when the $g_\chi=1$. Especially, when $m_{H_2}=400$ GeV, the contribution from the on-shell $H_2$ is still dominating even with the small kinematic phase space. This is in contrast to the Fig.~\ref{mdd} where the decay width of $H_2$ is much wider rendering the disappearance of the $H_2$ peak. 
We note that differences in the distributions of $m_{DD}$ between the FDM and VDM only occur in the off-shell $H_2$ processes. Otherwise, it is simply the on-shell $H_2$ production with subsequent invisible decay, which leaves no information of DM spin in the visible products. This explains why the $m_{DD}$ distributions for FDM and VDM almost overlap when $H_2$ is light, while the difference becomes visible in the region $m_{H_2} \gtrsim 300$ GeV where the off-shell contribution is sizable. 
For $g_\chi=10$ which is close to the perturbative limit, the decay width of $H_2$ is so wide that the off-shell $H_2$ contribution is important when $m_{H_2} \lesssim 200$ GeV and is dominant when $m_{H_2} \gtrsim 300$ GeV. Then, it is possible to distinguish the FDM against VDM in the full range of $m_{H_2}$. From the right panel of Fig.~\ref{mddg110}, we can also see that $m_{DD}$ distributions for FDM (VDM) with $m_{H_2} \gtrsim 300$ GeV are almost identical, because the signal events are occupying the lower side of the off-shell $H_2$ propagator irrespective of the $H_2$ mass and decay width. 

The signals are searched with the same method as used for benchmark points with $g_\chi=3$. We will only discuss the hadronic channel, since we have shown that it has better sensitivity than the leptonic channel. We first list the production cross sections of the benchmark points in the hadronic channel and the corresponding preselection efficiencies in Table~\ref{Fg110}. 
Compared to Table~\ref{Fg3}, we can find that all benchmark points in the FDM have similar total production rate when the $H_2$ is relatively light. While for $m_{H_2} \gtrsim 300$ GeV, the production cross section increases with the coupling $g_\chi$. The increase is more dramatic for heavier $H_2$. Eventually, the signal production cross sections are approaching to the same value when $g_\chi$ is close to the perturbative limit due to the dominance 
of the off-shell $H_2$ contribution. The preselection efficiency for most of the benchmark points are similar, i.e. between 0.7-0.8, except for the FDM400 with $g_\chi=1$.  For this benchmark point with $g_\chi=1$, the final state 
particles are a $Z$ boson ($m_{Z}=91.2$ GeV) and an almost on-shell $H_2$ ($m_{H_2} =400$ GeV), rendering the kinetic energy of final states quite small, $E^{\text{kin}} \sim \mathcal{O}(10)$ GeV. The preselection condition $E^{\text{miss}}_T >50$ GeV can cut out a large number of events.  

\begin{table}[htb]
\begin{center}
\begin{tabular}{|c|c||c|c|c|c|c|} \hline
\multicolumn{2}{|c||}{}  & FDM200 & FDM300 & FDM400 & FDM500\\ \hline \hline
\multirow{ 6}{*}{$g_\chi=1$} & $\sigma^0$ [fb] & 1.73 & 0.85 & 0.15 & 0.031   \\  \cline{2-6}
& $\epsilon^{\text{pre}}$ & 0.799 & 0.700 &  0.334 & 0.686   \\ \cline{2-6}
& BDT  & 0.3391 & 0.2383 & 0.0564 & 0.2402  \\ \cline{2-6}
& $N_S$/1000 fb$^{-1}$ & 774 & 374.6 & 38.1 & 10.8  \\ \cline{2-6}
& $N_B$/1000 fb$^{-1}$ & 1922.2 & 6348.9 & 31910.6 & 9130.4  \\ \cline{2-6}
& $N_S/\sqrt{N_S+N_B}$ & 14.9 & 4.6 & 0.213 & 0.113  \\ \hline \hline
\multirow{ 6}{*}{$g_\chi=10$} & $\sigma^0$ [fb] & 1.78 & 1.88 & 1.80 & 1.76  \\  \cline{2-6}
& $\epsilon^{\text{pre}}$ & 0.776 & 0.735 & 0.731 & 0.738  \\ \cline{2-6}
& BDT  & 0.2931 & 0.2610 & 0.2706 & 0.2816  \\ \cline{2-6}
& $N_S$/1000 fb$^{-1}$ & 762.8 & 755 & 706.6 & 697  \\ \cline{2-6}
& $N_B$/1000 fb$^{-1}$ & 5105 & 7416 & 7293 & 7194  \\ \cline{2-6}
& $N_S/\sqrt{N_S+N_B}$ & 9.96 & 8.35 & 7.9 & 7.8 \\ \hline \hline
\end{tabular}
\caption{\label{Fg110} The meaning of each row is the same as in Table~\ref{Fg3}, but the $g_\chi$ of benchmark points are changed to 1 and 10 for the upper half and lower half of the table, respectively.  }
\end{center}
\end{table}

The same BDT method that has been used in subsection~\ref{sec:fdmg3} for benchmark points with $g_\chi=3$ is also adopted here. The BDT is trained on the preselected events of each benchmark point with given $g_\chi$ and $m_{H_2}$ in the FDM model and the SM background. A cut on the BDT responses of signal and background can be applied later to improve the signal significance. 
The BDT cut for each benchmark point that maximizes the signal significance $(N_S/\sqrt{N_S+N_B})$ is given in  Table~\ref{Fg110}. We can find that at the ILC with $\sqrt{s}=500$ GeV and $\mathcal{L}=1000$ fb$^{-1}$, for $g_\chi=1$, only the benchmark points FDM200 and FDM300 can be discovered at more than 3-$\sigma$ level while for $g_\chi=10$, all of the benchmark points can be discovered with signal significance great than $\sim 8$-$\sigma$. 

The production cross sections and the preselection efficiencies of benchmark points in SDM and VDM models corresponding to those in FDM model with $g_\chi=1$ and $g_\chi=10$ are listed in Tables~\ref{SVg1} and 
\ref{SVg10}, respectively. 
For the case of $g_\chi=10$, the benchmark points in VDM model has much larger (smaller) production cross section than those in FDM model when the $H_2$ is light (heavy). So that it is possible to distinguish FDM and VDM even by using the production rates of signal alone. 
The number of signal events for each benchmark point in the SDM and VDM model after applying the BDT cut as well their significance level $\mathcal{S}$ of spin discrimination are calculated with the same strategy as introduced in subsection~\ref{sec:fdmg3}.
In the case of $g_\chi=1$, we can see in Table~\ref{SVg1} that only benchmark points SDM200 model can be distinguished from FDM model with $\mathcal{S} > 3$, while it is impossible to discriminate the FDM benchmark points against the VDM benchmark points. 
When the $g_\chi$ is close to the perturbative limit, the spin discrimination is quite promising as given in Table~\ref{SVg10}. 
The DM spin of our benchmark points with $H_2$ in the full mass region of interests can be identified with high significance level. Owning to the considerable difference in the production rate between the FDM and VDM, the VDM has better discriminating power against FDM than the SDM.

\begin{table}[htb]
\begin{center}
\begin{tabular}{|c||c|c|c|c|c|} \hline
  & SDM200 & SDM300 & VDM200 & VDM300\\ \hline
$\sigma^0$ [fb] &2.90 & 7.20 & 1.74 & 0.84  \\  \hline
$\epsilon^{\text{pre}}$ & 0.787 & 0.787 &  0.803 & 0.697  \\ \hline 
$N_S$/1000 fb$^{-1}$ & 774.0 & 374.6 & 777.6 & 363.5 \\ \hline
$\mathcal{S}$ & 3.36 & 1.14 & - & - \\ \hline
\end{tabular}
\caption{\label{SVg1} The meaning of each row is the same as in Table~\ref{SVg3}, but the benchmark points 
have been changed to those which are corresponding to benchmark points with $g_\chi=1$ in FDM model.  }
\end{center}
\end{table}

\begin{table}[htb]
\begin{center}
\begin{tabular}{|c||c|c|c|c|c|c|c|c|} \hline
  & SDM200 & SDM300 & SDM400 & SDM500 & VDM200 & VDM300 & VDM400 & VDM500\\ \hline
$\sigma^0$ [fb] & 1.82 & 1.58 & 1.48 & 1.47 & 2.08 & 1.77 & 1.02 & 0.643  \\  \hline
$\epsilon^{\text{pre}}$ & 0.7875 & 0.7875 & 0.7875 & 0.7875 & 0.774 & 0.725 & 0.720 & 0.718  \\ \hline 
$N_S$/1000 fb$^{-1}$ & 762.8 & 755.0 & 706.6 & 697.0 & 848.3 & 633.2 & 360.4 & 228 \\ \hline
$\mathcal{S}$ & 3.4 & 4.6 & 4.0 & 3.9 & 10.0 & 8.4 & 7.9 & 7.8 \\ \hline
\end{tabular}
\caption{\label{SVg10} The meaning of each row is the same as in Table~\ref{SVg3}, but the benchmark points 
have been changed to those which are corresponding to benchmark points with $g_\chi=10$ in FDM model.}
\end{center}
\end{table}

\section{Conclusion}
\label{sec:concl}

In this paper, we have considered DM discovery prospect and its spin discrimination  at the ILC in the theoretical 
framework of gauge invariant and renormalizable Higgs portal DM models for the first time. 
The gauge invariances of the FDM model and the VDM model require another new scalar field (in addition to the SM Higgs boson) that mediates the DM 
and SM particles interaction, while the gauge invariant SDM model only needs one medatior, the SM Higgs boson. 

Taking the FDM model with $g_\chi=3$ as a benchmark scenario, we study the discovery prospects and spin discriminating powers of both its hadronic channel and leptonic channel at the ILC with $\sqrt{s}=500$ GeV and $\mathcal{L}=1000$ fb$^{-1}$. 
In the hadronic channel, we first employ the BDT method with input of a few discriminative kinematic variables such as the DM pair invariant mass $m_{DD}$ and the azimuthal angular separation between the missing transverse momentum and the closer jet $\Delta \phi^{\min}$ to improve the signal sensitivity. 
We find the benchmark points with $m_{H_2} \lesssim 300$ GeV can be probed at more than 3-$\sigma$ level. For those discoverable benchmark points in the FDM model, the spin discriminating against SDM can be made with $\gtrsim$3-$\sigma$ level, due to the intrinsic difference between the FDM model and the SDM model, i.e. the FDM model contains two mediators while the SDM model only gets one. 
However, the spin discriminating against VDM is almost impossible, with the significance level below one for all discoverable benchmark points. 
The leptonic channel is also considered with the similar strategy. We find that the leptonic channel has worse 
discovery potential than the hadronic channel. Only benchmark points of FDM model with the mediator mass 
$m_{H_2} \lesssim 200$ GeV is discoverable. As with the hadronic channel, the spin discrimination between FDM and SDM can be made while it is quite difficult to distinguish FDM and VDM. 

We also survey the discovery and the spin characterizing prospects of the benchmark points in the FDM model 
with varying $g_\chi$. 
Choosing smaller $g_\chi$ does not reduce the DM production cross section in benchmark points with small $m_{H_2}$ much as long as the $H_2 \to \chi \chi$ branching is dominating. 
Furthermore, the smaller $g_\chi$ which gives narrower decay width of $H_2$ will increase the difference between the $m_{\chi\chi}$ distributions of the FDM and the SDM models. 
Thus benchmark points with $g_\chi=1$ even have better signal significances and spin discriminating powers than those with $g_\chi=3$. 
As for benchmark points with $g_\chi$ approaching the perturbative limit, the off-shell  $H_2$ contribution becomes quite important, leading to the increased production rate especially for those with heavy $H_2$. 
We find that the benchmark points with $H_2$ in the full mass region of interest are discoverable. The spin discriminating against both the SDM and VDM are quite promising. 

It should be noted that for FDM/VDM comparison throughout the work, the benchmark points of VDM are chosen such that the decay widths of $H_2$ are kept the same as the ones in the FDM model. This can be possible provided that the decay width of $H_2$ can be measured elsewhere. Then, the normalization of $m_{DD}$ distribution become an important handle for FDM and VDM discrimination. 
We also considered the FDM/VDM comparisons without the information of normalization and find the discrimiantions are impossible except for the cases of $g_\chi =10$. The $\mathcal{S}$ calculated from Eq.~\ref{eq:chi2} are 1.07, 1.24, 1.56 and 1.48 for FDM200/VDM200, FDM300/VDM300,  FDM400/VDM400 and FDM500/VDM500, respectively. 

\section*{Acknowledgement}

We are grateful to Bhaskar Dutta and Tathagata Ghosh for discussions on the related issues.
This work is supported in part by National Research Foundation of Korea (NRF) Research Grant NRF-2015R1A2A1A05001869 (PK, JL), and by the NRF grant funded by the Korea government (MSIP) 
(No. 2009-0083526) through Korea Neutrino Research Center at Seoul National University (PK). 
TK is partially supported by DOE Grant DE-SC0010813. TK is also supported in part by Qatar National Research Fund under project NPRP 9-328-1-066.

\phantomsection
\addcontentsline{toc}{section}{References}
\bibliographystyle{jhep}
\bibliography{DMspinILC}
\end{document}